%% file: fairgames.tex
\documentclass[journal]{IEEEtran}
\ifCLASSINFOpdf
\else
\fi
\input{preamble}

\begin{document}
%
\title{Offsetting Unequal Competition through RL-assisted Incentive Schemes}
%
%
%
\author{Paramita~Koley\textsuperscript{\textsection},
        Aurghya~Maiti\textsuperscript{\textsection},
        Sourangshu Bhattacharya,
        and Niloy Ganguly
\thanks{Paramita Koley, Sourangshu Bhattacharya and Niloy Ganguly are with the Department
of Computer Science and Engineering, Indian Institute of Technology Kharagpur,
Kharagpur-721302, India 
(e-mail: paramita.koley@iitkgp.ac.in; 
sourangshu@cse.iitkgp.ac.in;
niloy@cse.iitkgp.ac.in).}
\thanks{Aurghya Maiti is with Adobe Inc. (e-mail: aurghya.kgp@gmail.com)}
}
%
%

\if{0}
\markboth{Journal of \LaTeX\ Class Files,~Vol.~14, No.~8, August~2015}%
{Shell \MakeLowercase{\textit{et al.}}: Bare Demo of IEEEtran.cls for IEEE Journals}
%
\fi



\maketitle

\begingroup\renewcommand\thefootnote{\textsection}
\footnotetext{Equal contribution}
\endgroup

\input{000abstract}

\if{0}
\begin{IEEEkeywords}
IEEE, IEEEtran, journal, \LaTeX, paper, template.
\end{IEEEkeywords}
\fi

%
\IEEEpeerreviewmaketitle

\input{010intro}

\input{011Related_Work}
\input{020Problem}

\input{021marlexpt}

\input{030Algorithm}

\input{050Dynamic}
\input{050Conclusion}
\section*{Acknowledgment}

The authors would like to thank Intel Corporation for supporting this research.

\if{0}
\ifCLASSOPTIONcaptionsoff
  \newpage
\fi
\fi



\bibliographystyle{IEEEtran}
\bibliography{IEEEabrv,fairgames}
\if{0}

\fi
%
\vspace{-12mm}
\begin{IEEEbiographynophoto}{Paramita Koley}
	did her B.E. from IIEST Shibpur in 2010, M.E. from IISc Bangalore in 2013 and currently pursuing Ph.D. in IIT Kharagpur. Her research interests lie in learning with temporal point processes and various applications of reinforcement learning. 
\end{IEEEbiographynophoto}
\vspace{-10mm}
\begin{IEEEbiographynophoto}{Aurghya Maiti}
	received his B.Tech in computer science from IIT Kharagpur in 2020 and currently working in Adobe Inc, Bangalore. His research interests lie in reinforcement learning and multi-armed bandit problems. 
\end{IEEEbiographynophoto}
\vspace{-10mm}
\begin{IEEEbiographynophoto}{Sourangshu Bhattacharya}
 is an Assistant Professor in Dept. of Computer Science and Engineering at IIT Kharagpur. He has done his Ph.D. in Computer Science from IISc, Bangalore. Before joining IIT Kharagpur, he was a Scientist at Yahoo! Labs Bangalore, and a visiting scholar at the Helsinki University of Technology. He is a member of the ACM, and has been in the organising committees of ACM CODS-COMAD, SIAM Data Mining Conference etc and more than 20 publications in top international conferences and journals. He is broadly interested in machine learning, with specific interests in deep multi-task learning, learning with temporal point processes, network representation learning, and scalable and distributed machine learning.
\end{IEEEbiographynophoto}
\vspace{-10mm}
\begin{IEEEbiographynophoto}{Niloy Ganguly}
is a Professor in the Dept. of Computer Science and Engineering at IIT Kharagpur and a Fellow of Indian Academy of Engineering. He spent 2 years as a Research Scientist in Technical University, Dresden, before joining IIT Kharagpur in 2005. He has done his Btech from IIT Kharagpur and his Phd from IIEST, Shibpur. His research interests lie primarily in Social Computing, Machine Learning, and Network Science. He has several publications in reputed international venues such as NeurIPS, KDD, ICDM, IJCAI, WWW, CSCW, EMNLP, CHI, IEEE and ACM Transaction etc. 
\end{IEEEbiographynophoto}
\vfill

\if{0}
\begin{IEEEbiographynophoto}[{\includegraphics[width=1in,height=1.25in,clip,keepaspectratio]{AM.jpg}}]{Aurghya Maiti}
received the B.S. degree in computer science from the University of Arkansas, Fayetteville, AR, USA, in 2016, where he is currently pursuing the Ph.D. degree. His research focus is the development of an intelligent cyber argumentation platform and modeling and analyzing various phenomena in cyber argumentation.
\end{IEEEbiographynophoto}
\fi



\end{document}

%% file: preamble.tex
\usepackage{booktabs}       
\usepackage{amsfonts}       
\usepackage{nicefrac}       
\usepackage{microtype}      
\usepackage{amsmath}
\usepackage{cleveref}
\usepackage{color}
\usepackage{graphicx}
\graphicspath{ {./Results/} }
\usepackage{multirow}
\usepackage[caption=false]{subfig}
\usepackage{placeins}
\usepackage{float}
\usepackage[linesnumbered,ruled]{algorithm2e}
\usepackage{wrapfig}
\usepackage{lipsum}

\newcommand{\tT}{\mathcal{T}}
\newcommand{\aA}{\mathcal{A}}

\newcommand{\RR}{\mathcal{R}}
\newcommand{\our}{C-MADDPG}
\newcommand{\ourgame}{Touch-Mark}

\newcommand{\bsln}{MADDPG}

\newcommand{\niloy}[1]{\textcolor{red}{NG: #1}}

\crefname{section}{Section}{Sections}

\newcommand{\Acal}{\mathcal{A}}

\newcommand{\Ocal}{\mathcal{O}}
\newcommand{\Pcal}{\mathcal{P}}

\newcommand{\Rcal}{\mathcal{R}}
\newcommand{\Scal}{{\mathcal{S}}}

\newcommand{\PP}{\mathbb{P}} 


%% file: 000abstract.tex
\begin{abstract}
This paper investigates the  dynamics of competition among organizations with unequal expertise.
Multi-agent reinforcement learning has been used to simulate  and understand the impact of 
various incentive schemes designed to offset such inequality.
We design \ourgame{,} a game based on 
well-known multi-agent-particle-environment, where two teams 
(weak, strong) with unequal but changing skill levels compete against each other. 
For training such a game, we propose a novel controller assisted multi-agent reinforcement learning algorithm \our\, which empowers each agent with an ensemble of policies along with a supervised controller that by selectively partitioning the sample space, 
triggers  intelligent role division among the teammates. 
Using \our\  as an underlying framework, we propose an incentive scheme for the weak team such that the 
final rewards of both teams become the same. 
We find that  in spite of the incentive, the final reward of the weak team falls short of the strong team. 
On inspecting, we realize that an overall incentive scheme for the weak team does not incentivize the 
{\em weaker} agents within that team to learn and improve. To offset this,  we now specially incentivize the {\em weaker player} 
to learn and as a result, observe that the weak team 
beyond an  initial phase performs at par with the stronger team. The final goal of the paper has been to formulate a 
dynamic incentive scheme that continuously balances the reward of the two teams. 
This is achieved by devising an  incentive scheme enriched with an RL agent which takes minimum information from the environment.
\end{abstract}

%% file: 010intro.tex
\section{Introduction}
\label{sec:intro}

Society has evolved many mechanisms to offset  
inequality in real-life where often unequal individuals/teams/organizations have to compete against each other, namely, through  affirmative action~\cite{foster1992economic,austen2006redistribution}, special incentives~\cite{weisskopf2004impact,ali2015prevent,mukherjee2017conditional}, tax breaks~\cite{suarez2000does,alexander2009measuring,mckinnon2012firms}, compensation~\cite{perry2001pay,chan2014compensation}, subsidies~\cite{schwartz1999government,amegashie2006economics} etc.
While there are plenty of evidences that these measures help the weaker team, controversy persists around the implementation detail.  
For example, in the economics literature, it is well recognized that the use of subsidy by the government in a competitive market can improve welfare,
help domestic industry to compete against international counterparts, correct a market failure, bring social and private costs into alignment, to name only a few~\cite{amegashie2006economics,danglun2007empirical,zhao2014review,juriaith2014economics,giupponi2018subsidizing} and in the process successfully eliminate the very premise which has led to the introduction of subsidy. 

However, side by side, there are a series of works around
perverse subsidy~\cite{robin2003perverse,mackintosh2006perverse,si2006perverse,srinivasan2009subsidy,stephan2012perverse,chang2018lesson} 
which argue that when the subsidy is not directed towards
the right person or event, 
it may inflict various adverse effects on the economy, like higher tax, inefficient transfer of fiscal resources, or supply-side distortions, among many other possibilities.
Hence, it is safe to argue that the design of incentive schemes to neutralize the disadvantage suffered by a weak team is a non-trivial exercise. 
Moreover, given the nature of the problem, it is very difficult to  continuously monitor agents' responses and accordingly design  a dynamic incentive mechanism.

This paper looks into this problem by considering a simple {\bf  multi-agent reinforcement learning (MARL) framework} which 
allows us to monitor the response of agents towards incentive schemes and dynamically adjust them in real-time.
To the best of our knowledge, there is no study on continuous monitoring of agents' responses towards incentive.
The framework   has two major components: (1) a  multi-agent two-team game, 
called \textbf{\ourgame} which simulates competition and cooperation among the agents, and accommodates 
varying levels of agent \textbf{skills}; and 
(2) a controller assisted multi-agent reinforcement learning algorithm, called \textbf{\our{.}} 
\our\ builds over \bsln\ ~\cite{lowe2017multi}, however, unlike \bsln{,}
allows  efficient learning of an ensemble of agent policies and  provides a controller. It facilitates dynamic switching among the policies based on the situation, thus leading to the two agents  of a team taking up complementary  roles to ensure win.
We further postulate that  experiences from different roles impact the future skill level of an individual agent differently.

Using the above-described innovative framework, we study various static and dynamic incentive schemes
considering  two teams with unequal skill levels.
We find that the incentive given to the weaker team is effective and sustainable only when we direct targeted  incentives towards
the weaker players within the weaker team.
Based upon this finding, we design a dynamic incentive scheme that starts with a high value of additional  reward for the weaker team 
and gradually decreases  as the weaker team learns the winning policy and progressively becomes stronger.
There are several design issues related to the development of such a dynamic incentive mechanism. Most importantly, in real life situations,  it may not 
be possible to measure  certain performance related parameters dynamically. We tackle this issue by designing an RL agent that helps to dynamically 
predict the non-measurable parameters and design an  effective incentive scheme.
To summarize, this paper provides a simple setup  featuring a handful of characteristics from real-world team competitions, allowing us to study the effects of various incentives in a real-world competitive setting.

\textbf{Contributions:} To summarize, the main contributions of this paper are: (1) We initiate the study of agent and team performance in the setting of unequal and changing skill levels, through a novel game - \ourgame. (2) We study mechanisms of offsetting unequal competition through individual and team rewards, which can also be learned using RL. (3) We propose \our{,} which learns a dynamic role-based policy ensemble, for faster learning of agent policies and overall smaller simulation time.

\if{0}
Organizations often provide incentives to agents in order to achieve performance objectives, e.g. companies giving performance linked pays \footnote{https://en.wikipedia.org/wiki/Incentive\_program}, or various National Research agencies providing research funding \cite{stephan2012perverse}. However, sometimes such incentives do not benefit the agent \cite{stephan2012perverse} or leads to fairness concerns \cite{abc}. Moreover, the agents in the organisation may adopt roles based on their incentives (or rewards), and their own skill levels. For example, an individual with low communication skills may take on a back office job, than a customer facing role. Skill levels of agents may change over time, making them more suitable for other roles. However, the incentive scheme may still be tuned towards the previous role, giving them an unfair disadvantage. For example, the employee could be given incentive based on the number of documents processed, and not on customer satisfaction levels.
In this paper, we study the problem of automatically designing dynamic and fair incentive structures.
\fi

%% file: 011Related_Work.tex
\section{Related Work}
\label{sec:related_work}

Multi-agent reinforcement learning is a long-studied problem in various settings, namely learning joint strategy for cooperative tasks~\cite{guestrin2002coordinated,rangwala2019learning,wang2020cooperation, yang2020q}, optimal play in competitive setting~\cite{littman1994markov}, learning robust policies under model uncertainty~\cite{zhang2020robust} etc.
A very common and popular approach is the recent actor-critic framework consisting of centralized training with decentralized execution
~\cite{gupta2017cooperative}. MADDPG~\cite{lowe2017multi}, a multi-agent extension of deep deterministic policy gradient~\cite{lillicrap2015continuous}, is one such stable and popular algorithm, 
There are many follow-up works on actor-critic based MARL algorithms, namely multi-actor-attension-critic (MAAC) ~\cite{iqbal2018actor} for introducing attention, ~\cite{qu2020scalable} for improving scalability, ~\cite{christianos2020shared} for sharing experience, ~\cite{zhou2020learning} for credit assignment problem, ~\cite{mahajan2021tesseract} for tensorizing the critics,  etc.
In our game setting, we find the MAAC performs similarly to \bsln\ while being much more computationally expensive and thereby we continue our experiments with \bsln\ only. However, our proposed framework can be adapted to other multi-agent reinforcement learning algorithms as well. 
Recently,~\cite{majumdar2020evolutionary} present an extension of MADDPG with separately learning individual and global goals in a population-based training paradigm. ~\cite{liu2021coach} tackles the problem of dynamic team composition in coach-player paradigm.
However, none of them explicitly address the setting of unequal competition, with focus on effects of incentives in offsetting the inequality.
Also, there is a series of works for role-oriented MARL for specialized domains like robo-soccer \cite{leottau2015study,urieli2011optimizing,ossmy2018variety}, football environment~\cite{roy2019promoting} showing how complex policies can be learnt by decomposing it into simpler sub-policies.
However, our primary focus is being to study and analyze the effect  of various incentives schemes 
on unequal agents; we test our hypothesis on \ourgame, a simple team-competitive game as it will be difficult to gain insights  on complex multi-player robo-soccer. 
Despite there exists recent work on the stability of mixed-strategy learning algorithms~\cite{mertikopoulos2019learning} or the conditions for the convergence to Nash equilibria in continuous action spaces~\cite{kamra2019deepfp}, we postpone such theoretical exploration to a future work.

The practice of applying intrinsic incentives in multi-agent reinforcement learning framework by a third party to manipulate the dynamics to obtain the desired outcome  is mostly found in various social dilemma games~\cite{mohamed2015variational,hughes2018inequity,jaques2019social,paquette2019no}. 

		~\cite{iqbal2019coordinated} employ intrinsic rewards for coordinated exploration. ~\cite{du2019liir} employ individual intrinsic rewards for stimulating diverse behavior among agents.  A form of \textit{general utility}, a non-linear function of state-action occupancy measure, has recently shown to be effective in practice via prioritizing exploration~\cite{mahajan2019maven,gupta2021uneven}, risk-sensitivity~\cite{qiu2020rmix}, and prior experience~\cite{le2017coordinated,lee2019improved}. ~\cite{zhang2021marl} establish theoretical guarantees of consistency and sample complexity for such general utility function.
In this work, our proposed dynamic incentive closely resembles the intrinsic rewards proposed in ~\cite{hughes2018inequity}, though the setting or motivation is quite different. 
~\cite{jiang2019learning} explores learning fair and stable strategies in resource sharing settings. 
Close to our line of work, ~\cite{zheng2021ai} present a machine-learning based economic simulation framework, where AI economist, a two-level, deep RL framework is used to train agents along with a social planner to provide a tractable solution to the optimal taxation problem, unlocking a computational learning-based approach to understanding economic policy. 
However, studying fair outcomes in competitive setting is still in its nascent stage.  

The work presented here is in line with the
design of fair incentive schemes, which is an important area in fair machine learning~\cite{calders2009building,zafar2015fairness,hardt2016equality,pleiss2017fairness}. 
More specifically, it adds to 
the recent studies, which  have focused  on the long-term effects of social groups on implementing fairness constraints. 
~\cite{hu2018short} devise a data-specific affirmative action strategy on US labor market, which in turn ensures that  the need for affirmative action diminishes as time progresses. 
~\cite{liu2018delayed} show delayed impact of existing fair classifiers on disadvantaged groups.  
They demonstrate that even in a one-step feedback model, common fairness criteria, in general, may 
not promote improvement over time. 
Similar sentiments are echoed in ~\cite{corbett2018measure} where they show classification parity can, perversely, harm the very groups they were designed to protect.
~\cite{mouzannar2019fair} address the important issue of maintaining demographic parity and quality.
~\cite{kannan2019downstream} discuss the relation between 
the constraint of equal opportunity in college admission and 
biases induced due to this during hiring by companies.
~\cite{jabbari2017fairness} build a reinforcement learning model which achieves near-optimality
subject to (exact) fairness or approximate-choice fairness.
Recently, there is a series of works at the intersection of incentive-based mechanism design and reinforcement learning \cite{zheng2020ai,brero2020reinforcement,zhang2021incentive}. 
Among theoretical works, \cite{brero2020reinforcement} investigate various theoretical aspects of the use of reinforcement learning for certain classes of indirect mechanisms whereas  
\cite{zhang2021incentive} propose incentive-aware PAC learning in the presence of strategic manipulation.
Our work closely resembles \cite{zheng2020ai} that build social planners for devising tax policies
in dynamic economies for effectively balancing economic equality and productivity, 
where the agents are trained through deep reinforcement learning.
The present work adds to the domain at the intersection of { \bf mechanism design, reinforcement learning and fair machine learning} 
by considering competition between unequal teams
and build RL agents for devising dynamic incentives for fair outcomes.


\if{0}
{\bf Related Work:} The work presented here is in line with the
design of fair incentive schemes which is an important area in Fair ML ~\cite{calders2009building,zafar2015fairness,hardt2016equality,pleiss2017fairness}. More specifically, it adds to 
the recent studies, which  have focused  on long term effects of implementing affirmative action. 
~\cite{hu2018short} devise a data-specific affirmative action strategy on US labor market which in turn ensures that  the need for affirmative action diminishes as time progresses. 
~\cite{liu2018delayed} show delayed impact of existing fair classifiers on disadvantaged groups.  
They demonstrate that even in a one-step feedback model, common fairness criteria, in general, may 
not promote improvement over time. Similar sentiments are echoed in ~\cite{corbett2018measure} where they show classification parity can, perversely, harm the very groups they were designed
to protect.
~\cite{mouzannar2019fair} address the important issue of maintaining demographic parity and quality.
~\cite{kannan2019downstream} discuss the relation between 
the constraint of equal opportunity in college admission and 
biases induced due to this during hiring by companies.
~\cite{jabbari2017fairness} build a reinforcement learning model which achieves near-optimality
subject to (exact) fairness or approximate-choice fairness.
However, hardly any work has discussed the issues pertaining to implementation of fairness constraints in a competitive setting or 
the need for long term incentive  after affirmative entrance. We believe the 
`data-independent' modeling technique which we introduce in this space is still largely unexplored. 

\fi
%
%


\if{0}
\niloy{We have to connect this paragraph}
Use of subsidy is a well studied topic in economics literature. It is well recognized that use of subsidy by government in competitive market can improve welfare,
correct a market failure, bringing social and private costs into alignment, etc~\cite{amegashie2006economics}. ~\cite{amegashie2006economics} also highlights that the practice of subsidizing a domestic industry, which is unable to compete with exports, is quite common for government, on a ground of learning-by-doing effects, from which the whole economy will be benefited. Also it considers subsidization is regarded as a more efficient option than other options like imposing tariff over exports. Though utmost care must be taken while devising subsidies, as improper subsidies can cause considerable damage to corresponding domain, an event, named as perverse subsidy~\cite{robin2003perverse,mackintosh2006perverse,si2006perverse,stephan2012perverse}. 
Besiddes having many positive sides, subsidies have some adverse effects on economy, like higher tax, inefficient transfer of fiscal resources etc
~\cite{tangermann2005organisation}.
In a similar tone, ~\cite{srinivasan2009subsidy,chang2018lesson} shows how an inappropriately designed long-term subsidy can result in  
supply-side distortions as well as being counter-productive, 
discouraging the use of subsidy for longer term or as a permanent solution.

\niloy{There are many works which show that long term subsidy is not good, only short term subsidy to face the competitive market is needed}

\fi

%% file: 020Problem.tex
\newcommand{\cC}{\mathcal{C}}
\newcommand{\bx}{\mathbf{x}}
\newcommand{\ba}{\mathbf{a}}
\newcommand{\br}{\mathbf{r}}
\newcommand{\cD}{\mathcal{D}}
\newcommand{\by}{\mathbf{y}}

\section{MARL under Team Competition}
\label{sec:problem_formulation}

We propose \ourgame{,} an episodic board game, 
built on the multi-particle environment (MPE) ~\cite{lowe2017multi}, which elicits both competitive and collaborative behavior among agents. In this game, we focus on three major aspects of social behaviour: (1) team competition, (2) the emergence of unequal roles, and (3) skill improvement. 
\ourgame~ is largely derived from \textsc{Keep-away}, a $2$ player competitive game introduced in~\cite{lowe2017multi}, where an agent and its adversary both are trying to reach the landmark while trying to push the opponent away from the landmark. 
For incorporating team competition, we increase the number of members in each team to at least $2$ and consequentially increase the number of landmarks to $2$ (encouraging diverse policies).
\textit{Reaching the landmark} and \textit{colliding with opponent}, an agent can adopt these two implicit roles  within a team as a policy of the gameplay. 

In this game, we also assign a {\bf skill} level to each agent, which improves over time. 
The improvement varies depending on which role the agent plays, to simulate the dynamics of how assigning more rewarding roles with more scope for self-improvement to more skilled ones leads to more inequality among the team members. 
Note that while analyzing the subsidy schemes in the following sections, we assign different initial {\bf skill} levels to different agents to model {\bf unequal competition}. 
While \ourgame\ is simple and efficient, it also incorporates all the features of team competition-based social interactions that are commonly seen in society. Hence, it can be used as a simulation platform for our studies on incentive schemes.
We exclude some of the more complex and popular team competition games, e.g. Google Football Environment~\cite{kurach2020google}, StarCraft 2~\cite{samvelyan2019starcraft} etc. because those are too heavy on computational resources as well as it is more complicated to analyze and differentiate the effects of various incentives.  
Next, we briefly describe the rules of \ourgame{.}

\subsection{\ourgame{:} Team Competition between Unequal Agents}
The game setting consists of two teams, each comprising of two agents. Each agent has its current position and velocity.
The game is played iteratively; 
in each \textit{episode}, two landmarks (which introduce diversity) are placed at random in a square board and each team tries to reach at least one of those landmarks earlier than  any member of the other team. The episode ends when an agent reaches a landmark. The winning agent's team  (ie. both team members) receives a large reward $r_l$, simultaneously $-r_l$ penalty is incurred to members of the opposite team. 
Additionally, each agent receives a small penalty,  $-r_d$, ($r_d ~<<~ r_l$)) which is proportional to its distance from 
the nearest target at every time step. The penalty encourages the agents to move towards the target,  thus accelerating the learning of policy. To stop agents from going out of the box, a small penalty is given for touching the boundary.
Moreover, an agent can collide with an agent of the opponent team to divert it from its path. This mechanism is introduced so that an agent 
has the option to stop an agent of the opponent team from reaching a landmark, thus facilitating the  fellow teammate to  reach a landmark first. This cooperative behavior results from the emergence of  different roles within a team.

In this game, each agent starts at a random position and zero velocity. Velocity of each agent $i$ can increase up to an upper bound (\texttt{max\_speed$_i$}). Each agent has a parameter (\texttt{max\_speed$_i$}), representing corresponding skill level, since it limits the speed at which the agent can move. 
In \ourgame{,} the \texttt{max\_speed$_i$} is increased at the end of each episode if the agent $i$ touches any landmark in that episode, representing skill upgrade at the end of success.
The rate of increase of \texttt{max\_speed$_i$} is proportional to its difference from a global speed limit \texttt{MAX\_SPEED}, i.e. higher the skill, slower is the rise in skill level.

Despite being apparently simple in nature, this game captures a few key aspects of the dynamics of competition. 
In real life, such a game resembles the setting where multiple organizations are competing for some common target and employees within organizations resemble the members within the team. 
In a similar analogy, touching the landmark mimics the target fulfillment by an organization. 
The provision for collision in \ourgame\ also is comparable to the situation where organizations put effort to outwit the competitor.

Most importantly, this game offers a skill improvement feature, i.e., an agent improves its skill (\texttt{max\_speed$_i$} in our case) on achieving some target which closely resembles the popular argument that government policy-makers put behind supporting an infant organization, the theory of learning-by-doing effect. Also the game assumes that one role is more important to attain the target, hence although the reward is equally divided among all the members, the skill improvement happens to one particular member. This mimics real life where, in a team, some may be doing desk job while others are doing field job and learning is much steeper in the field. In each team, 
the number of agents has been fixed at $2$. The difference in skills and role can both be captured through the two members and keeping the number small also  ensure sufficient interpretability of the effects of various incentive schemes introduced later. Increasing the number of players has disadvantages both ways,   it  becomes complicated/time-consuming 
and  it hinders from uncovering the impact of incentives on the ecosystem as well as on individual players.
\subsection{Learning Ensemble Policies}

We cast the problem of learning optimal policies for agents as a multi-agent reinforcement learning problem. We briefly define the formal setup for multi-agent reinforcement learning. 
We consider a multi-agent extension of Markov decision processes called partially observable Markov games~\cite{littman1994markov}.

A Markov game is characterized by the tuple $\langle \Scal,\Acal,\Pcal,\Rcal \rangle$, where $\Scal$ denotes the set of states, $\Acal$ denotes the set of actions for each of the $N$ agents.  Hence the joint space of actions becomes $\Acal^N$.
The state transition function maps every state and joint-action combination to a probability over future states. $\Pcal \colon \Scal \times \Acal^N \rightarrow \PP(\Scal)$. 
Reward function  $\Rcal = (r_1,\dots,r_N) \colon \Scal \times \Acal^N  \rightarrow \RR^N $ specifies the reward scheme ($r_i$) for each agent $i$.
The per-agent reward function allows modeling of team competitive games \cite{lowe2017multi}.
In the decentralized execution setting, each agent $i$ receives its own observation $o_i \in \Ocal$ which is a function of the common state $s$, but from the agents' point of view.  
Each agent $i$ learns a policy  $\pi_i \colon \Ocal \rightarrow \PP( \Acal )$, which is a mapping  from it's own observation to a distribution over it's action set $\Acal$.

\if{0}
The Markov game is characterized by 
the set of configurations $\mathcal{S}$, the set of actions $\mathcal{A}_1,\dots ,\mathcal{A}_N$,
and the set of observations $\mathcal{O}_1,\dots ,\mathcal{O}_N$ visible for agents. 
Each agent owns a policy $\pi_i$ to choose action from its action set, which produces next state according to transition function $\mathcal{\tau}$. 
Each agent $i$ obtains its rewards according to reward function $r_i$. 
\fi 

\begin{figure}[!t]
	\centering
	\subfloat{\includegraphics[width=.3\textwidth]{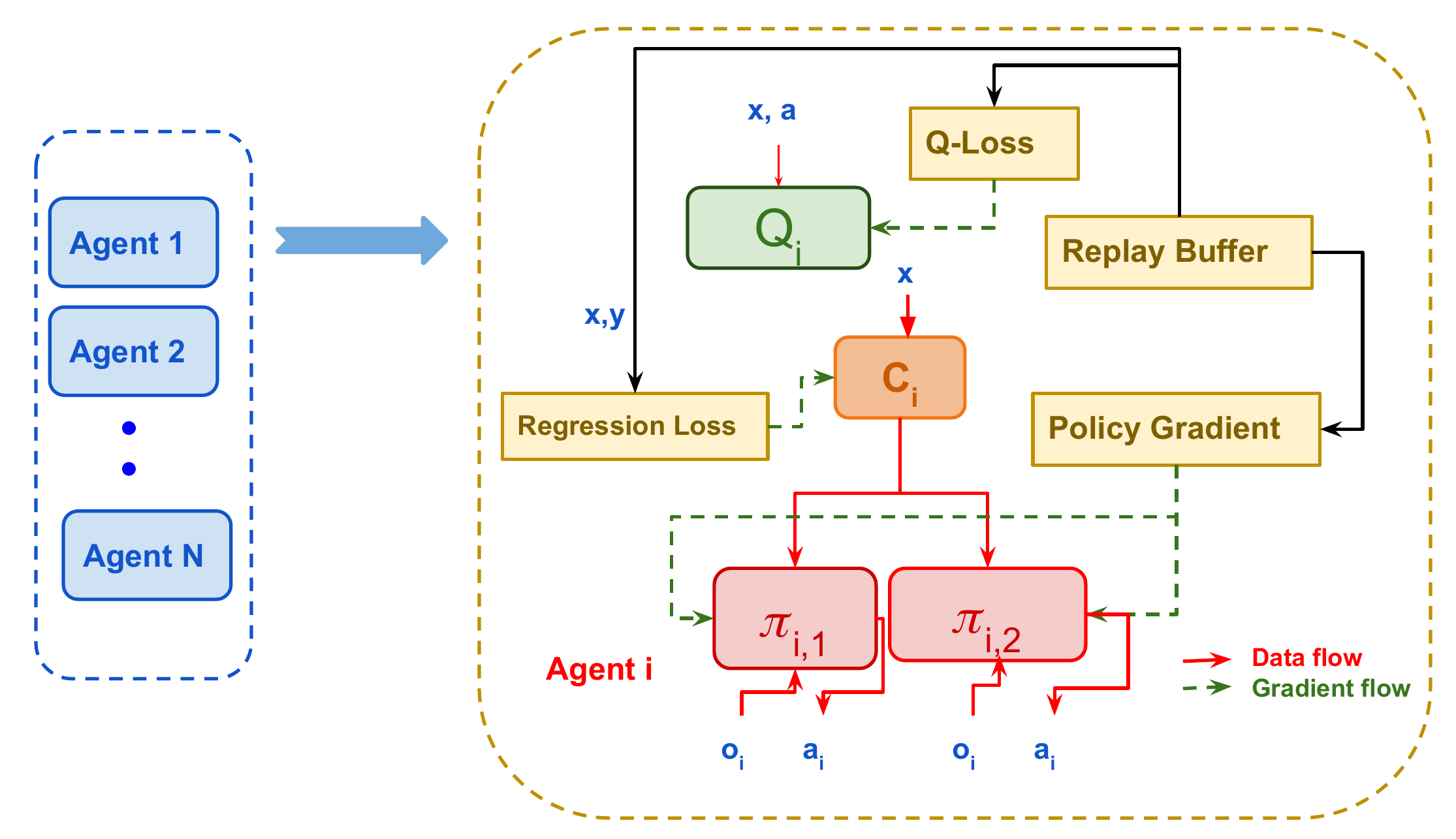}}
	\caption{Schematic diagram of \our}	
	\label{fig:flow-chart}
\end{figure}

\noindent {\bf \bsln{:}}
In above setup, we restrict ourselves to the well-known actor critic framework in the deterministic policy setting, where \textit{Multi-Agent Deep Deterministic Policy Gradient} (\bsln{)} \cite{lowe2017multi} is a popular state of the art method. In \bsln, each agent maintains its policy $\pi_i$, parameterized by $\theta_i$ and an approximation of the action-value function $Q_i$ parameterized by $\phi_i$ for $i=1,\dots , N$. 
In \bsln\, the policy $\pi_i$ 
is updated in a decentralized manner, whereas the action-value function (critic) $Q_i$ is trained in a centralized manner.

Given an experience replay buffer $\cD = \{ (\bx_t,\ba_t,\br_t,\bx'_t), t=1,\dots,T\}$, where $\bx = ( o_i )_{i=1}^{ N }$ are the current observations / states, $\ba = ( a_i )_{i=1}^{ N } $ are the current actions, $\br = ( r_i )_{i=1}^{ N }$ are the rewards, and $\bx'$ are the observations at next time; the gradient of cumulative reward function w.r.t. policy parameters $\theta_i$ can be computed as:
$
	\nabla _ { \theta _ { i } } J \left( \boldsymbol { \pi } _ { i } \right) = \mathbb { E } _ { \mathbf { x } , a \sim \mathcal { D } }
	\left[ \left. \nabla _ { \theta _ { i } } \boldsymbol { \pi } _ { i } \left( a _ { i } | o _ { i } \right) \nabla _ { a _ { i } } Q _ { i }  \left( \mathbf { x } , a _ { 1 } , \ldots , a _ { N } \right) \right| _ { a _ { i } = \boldsymbol { \pi } _ { i } \left( o _ { i } \right) } \right].
	$
The policy parameters are iteratively updated using the above gradients. The action-value functions $Q_i$ is learned by minimising w.r.t. $\phi_i$:
\begin{multline}\small
	\mathcal { L } \left( \phi _ { i } \right) = \mathbb { E } _ { \mathbf { x } , a , r , \mathbf { x } ^ { \prime } } \left[ \left( Q _ { i } \left( \mathbf { x } , a _ { 1 } , \ldots , a _ { N } \right) - y \right) ^ { 2 } \right] \ ;\\\ \quad y = r _ { i } + \left. \gamma Q _ { i } ^{ \prime }  \left( \mathbf { x } ^ { \prime } , a _ { 1 } ^ { \prime } , \ldots , a _ { N } ^ { \prime } \right) \right| _ { a _ { j } ^ { \prime } = \pi _ { j } ^ { \prime } \left( o _ { j } ^ { \prime } \right) }
	\label{eq:critic-update}
\end{multline}
where $\boldsymbol { \pi } ^ { \prime } = \left\{ \boldsymbol { \pi } _ { \theta _ { 1 } ^ { \prime } } , \ldots , \boldsymbol { \pi } _ { \theta _ { N } ^ { \prime } } \right\}$ is the set of delayed target policies parametrized  by $\theta_i^\prime$ and $Q_i^{\prime}$ is the delayed action value function parameterized by $\phi_i^{\prime}$. 

\begin{algorithm}
	\caption{ \our\ }
	\label{modifiedAlgo}
	\KwData{$M$: \#episodes, $T$: max episode length, $N$: \#agents, $\tau$: classfier update interval, $C$ : Constant}
	\textbf{Randomly initialize}: $\cC _i$ policy-classifier, $\phi_i$: $Q_i$ params, $\theta_i^j$: $\pi_i^j$ params, $\forall i\in [N],\ j\in \{1,2\}$ \;
	\For{$episode \gets 1$ \textbf{to} $M$}{
		\For{ $t \gets 1 $ \textbf{to } $T$ }{
		
		Select $\pi_i$ as exploration policy w.p. $e ^ {-episode/C}$\; \textbf{else} $\pi_i = \pi_i^j$ where $j = \cC_i(\bx_t)$ $\forall i$\;
		\tcc{Note that, $\cC_{i_1}(x_t)=\cC_{i_2}(x_t)$ if team$(i_1)$=team$(i_2)$}
			[Execute $a_i = \pi_i(o_i)$ and observe new state $\bx_{t+1}$ and reward $r_i(\bx_{t+1}) $] $\forall i$\;
			Calculate $\by_t = (y_i)_{i=1}^{N} $ using equation \ref{eq:y-update} \;
			Store $(\bx_t,\by_t,\ba_t,\br,\bx_{t+1})$ in replay buffer $\cD$\;
			\tcc{Update critic, policy}
			Sample $(\bx,\by,\ba,\br,\bx^{'})$ from $\cD$ \;
				Update $Q_i(\phi_i)$ using eqn \ref{eq:critic-update} $\forall i$\;
				Update $\pi_i^{j}(\theta_i^{j})$ using eqn \ref{eq:new-policy-update} $\forall i$, $\forall j$\;
				
		}
		Every $\tau$ episodes: update $ \mathcal{C}_i $ using gradient of cross-entropy loss $\nabla l(\cC_i(\bx),y_i)$ $\forall i$\;
	}		
\end{algorithm}

\noindent \textbf{\our{:}} 
A major problem with the \bsln\ algorithm is that each agent $i$ has only one set of policy parameters $\theta_i$ and it is difficult to learn complex policies using a single set of parameters. 
But \ourgame\ being a team game, the emergence of 
diverse behaviors within the team while trying to achieve a common goal may be efficient. 
To capture the diverse behavior, we develop \our, a controller assisted version of \bsln , 
which learns an ensemble of policies per agent and also maintains a controller per agent, that is trained in a supervised manner to switch between policies, based on the relative maximum critic function values of the teams.
To state our proposal formally, we propose to learn an ensemble of policies $\pi_i^j$, $j=1,\dots,J$ per agent. 
Here  we  consider $J=2$, $j=1$ corresponds to a  policy adopted at a configuration where the team is in  an 
advantageous position (we call it a winning policy)  and $j=2$ refers to  a losing policy. 

We propose a simple controller assisted training paradigm 
where at each timestep $t$ and for each agent $i$, we update (follow Algorithm~\ref{modifiedAlgo}): 
(1) two policies $\pi_i^j$, $j\in \{1,2\}$ which guide the next step of the agent (line $11$ of Algorithm~\ref{modifiedAlgo}) 
(2) a classifier $\cC_i(\bx)$ which takes the current observations of all agents $\bx$ and assigns a 
policy label $j\in \{1,2\}$ to the agents (line $13$ of Algorithm~\ref{modifiedAlgo}); 
The target policy labels, used in training the classifier $\cC_i$, are computed according to the following equation (line $13$ of Algorithm~\ref{modifiedAlgo}).
\begin{multline} 
\mbox{if}\left( \max_{j \in \mbox{team$_k$}} \{ Q_j(\bx_t,\ba_t) \}>  \max_{l\in \mbox{team$_m$}}\{Q_l(\bx_t,\ba_t)\} \right) \\
y_i = 1;\ \mbox{else}\ y_i = 2 \ \forall i \in \mbox{team$_k$}
\label{eq:y-update}
\end{multline}

 The equation ensures that  the classifier $\cC_i$  is the same for all the members of a team, a team adopts a winning policy ($j=1$) if any one member of the team possesses the highest critic value across all the agents and losing policy otherwise. 
Here, the underlying assumption is that $Q$ values of the agents are observable across the teams.

At each time-step $t$, each agent $i$ determines it's policy $j$ using the classifier $\cC_i$, (the construction of the classifier ensures that both members of a team choose the same policy) and 
then uses the policy $\pi_i^j$ to select its action $a_i(t)$ and receive reward $r_i(t)$ and $(\bx_t,\by_t,\ba_t,\br_t,\bx_{t+1})$ are stored in replay buffer.
The policies of the agents are updated using the following  equation at each timestep $t$, 

\begin{multline}\small	
	\nabla _ { \theta _ { i } ^{y_i}} J \left( \boldsymbol { \pi } _ { i } ^{y_i}  \right) = \mathbb { E } _ { \mathbf { x }, \by , \ba \sim \mathcal { D } }\\
\left[ \left. 
\nabla _ { \theta _ { i } ^{y_i} } \boldsymbol { \pi } _ { i }^{y_i} \left( a _ { i } | o _ { i } \right) \nabla _ { a _ { i } } Q _ { i }  \left( \mathbf { x } , a _ { 1 } , \ldots , a _ { N } \right) 
 \right| _ { a _ { i } = \boldsymbol { \pi } _ { i }^{y_i} \left( o _ { i } \right) } \right]
\label{eq:new-policy-update}
\end{multline}

Intuitively, the rationale behind role emergence in \our\ is as follows: the sample space is split based on the relative superiority of one team over another, and policies are trained over disjoint sample spaces. This modification allows \our\ to enjoy two-fold benefits: \our\ learns more focused, sub-goal oriented policies by systematically splitting the training sets as well as achieves the goal of complex policy by dynamic switching between the policies. Note that, we additionally require the opponent value functions as input, which are of the same size as the global states and actions. Hence, scalability is similar to CTDE methods like MADDPG for a small number of roles. Please refer to fig.~\ref{fig:flow-chart} for schematic representation of our algorithm.

%% file: 021marlexpt.tex
\begin{figure}[!t]
	\centering
	\subfloat[Team avg. rewards]{\includegraphics[width=.15\textwidth]{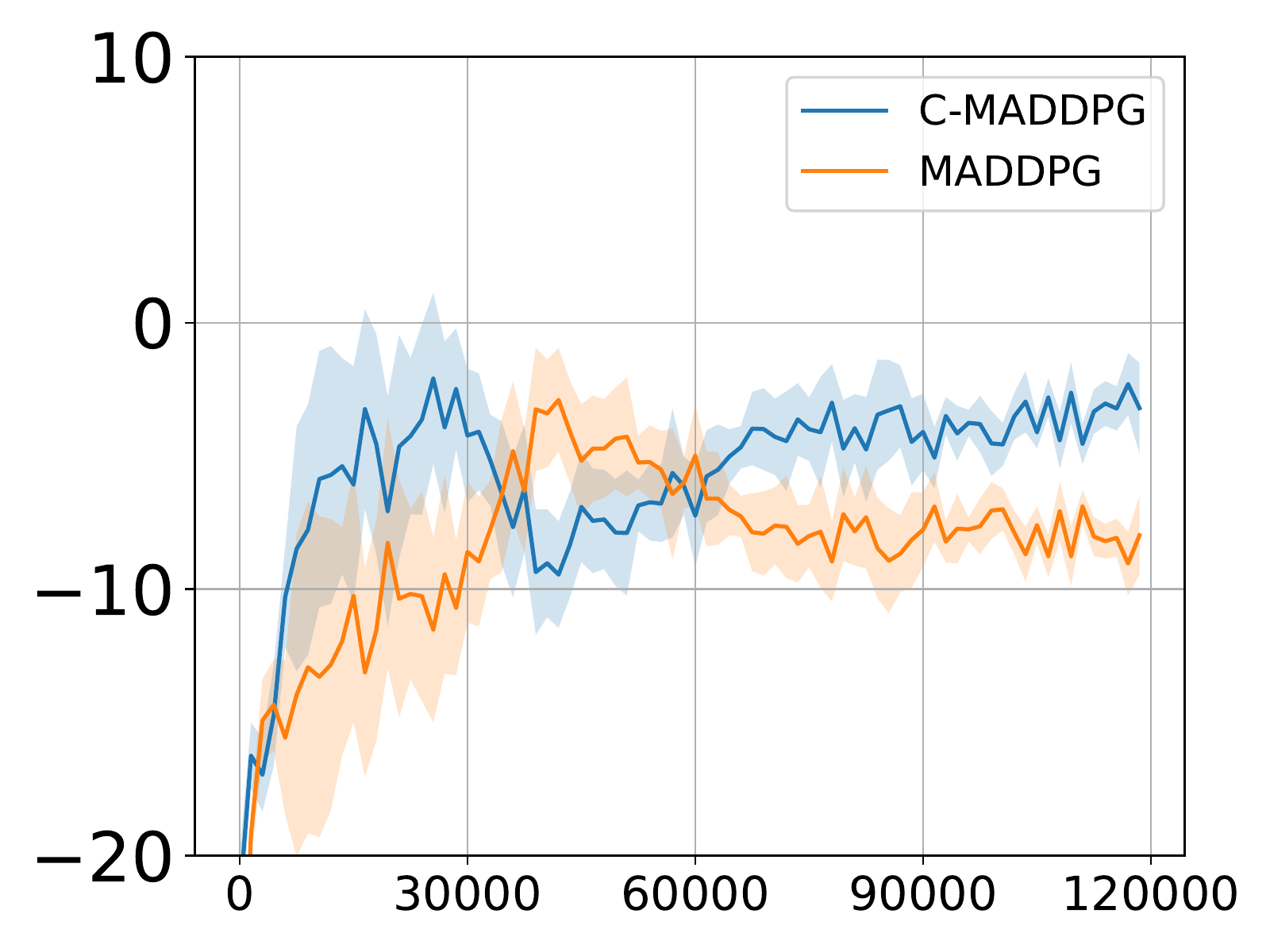}\label{fig:calibration-score}}\vspace{8mm}
	\subfloat[Agent landmark rate]{\includegraphics[width=.15\textwidth]{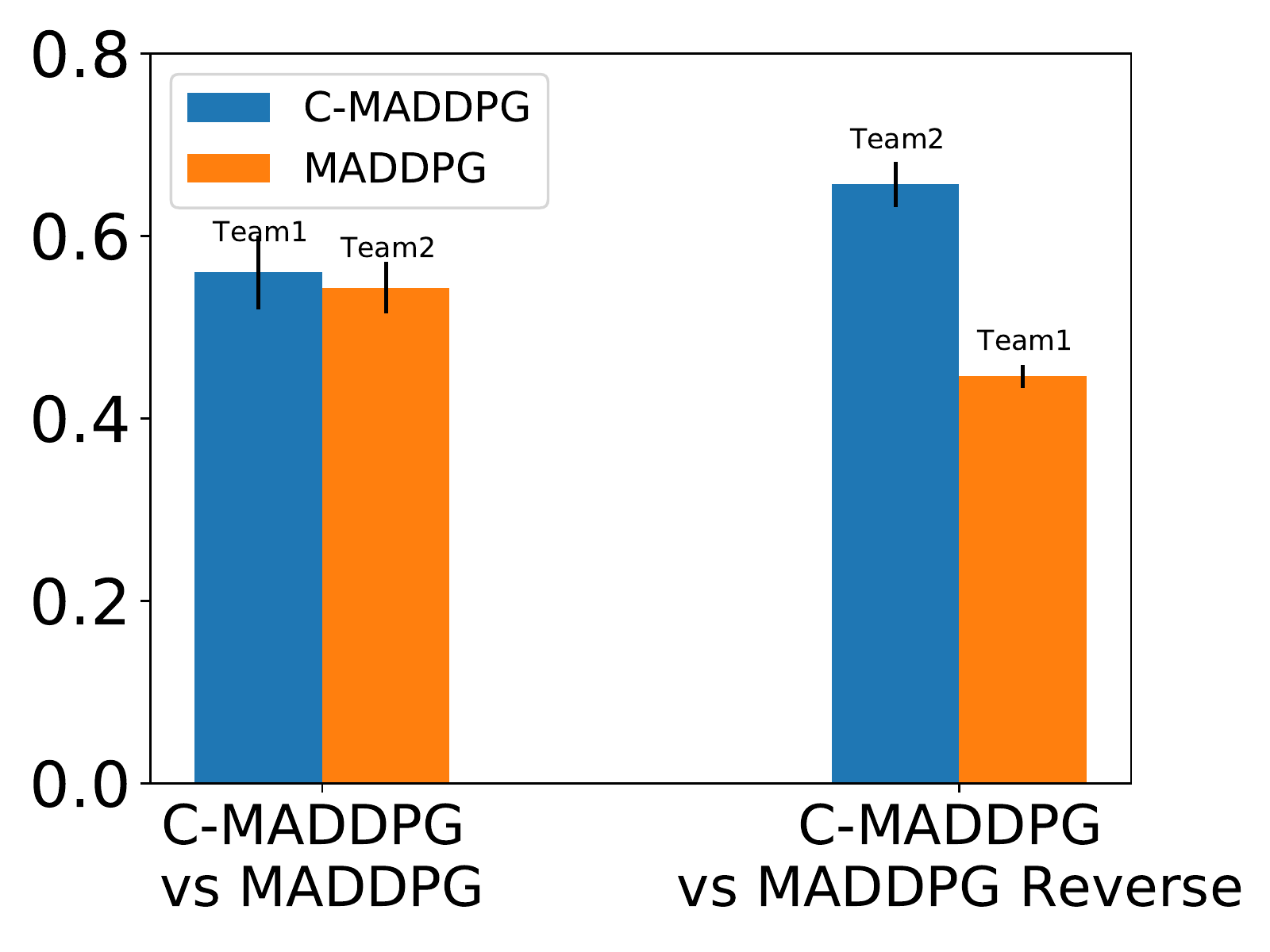}\label{fig:calibration-count}}	
	\caption{(a) Temporal evolution of team-wise average rewards. Experiments have been performed for six different seeds. 
		Shaded region denotes $95\%$ confidence region and (b) Team-wise landmark reaching rates for a set of initial configurations and its reverse.
	}	
	\label{fig:calibration}
\end{figure}

\begin{figure}[!t]
	\centering
	\subfloat[Forward position]{\includegraphics[width=.15\textwidth]{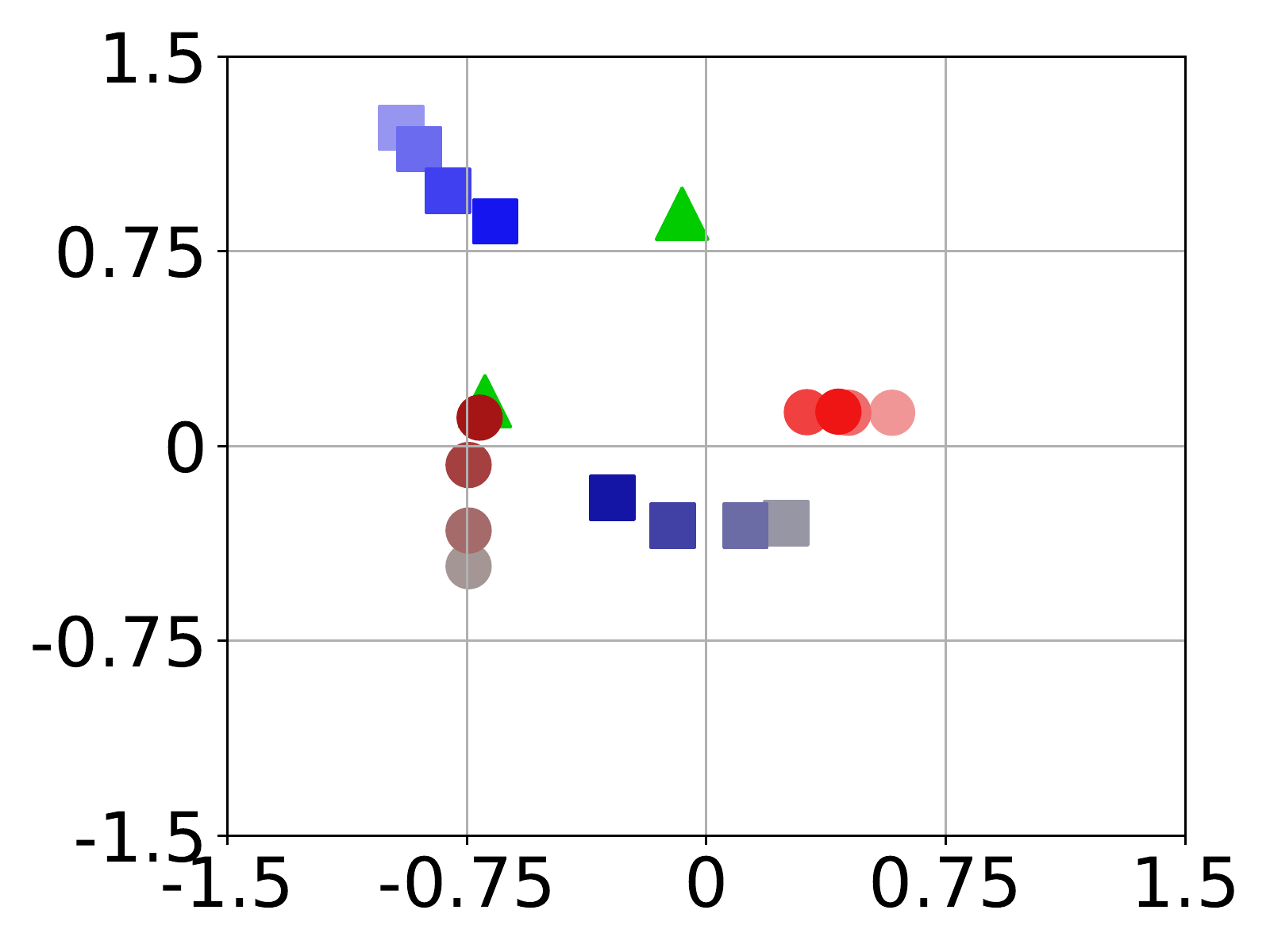}\label{fig:demo129}}\vspace{5mm}
	\subfloat[Flip position]{\includegraphics[width=.15\textwidth]{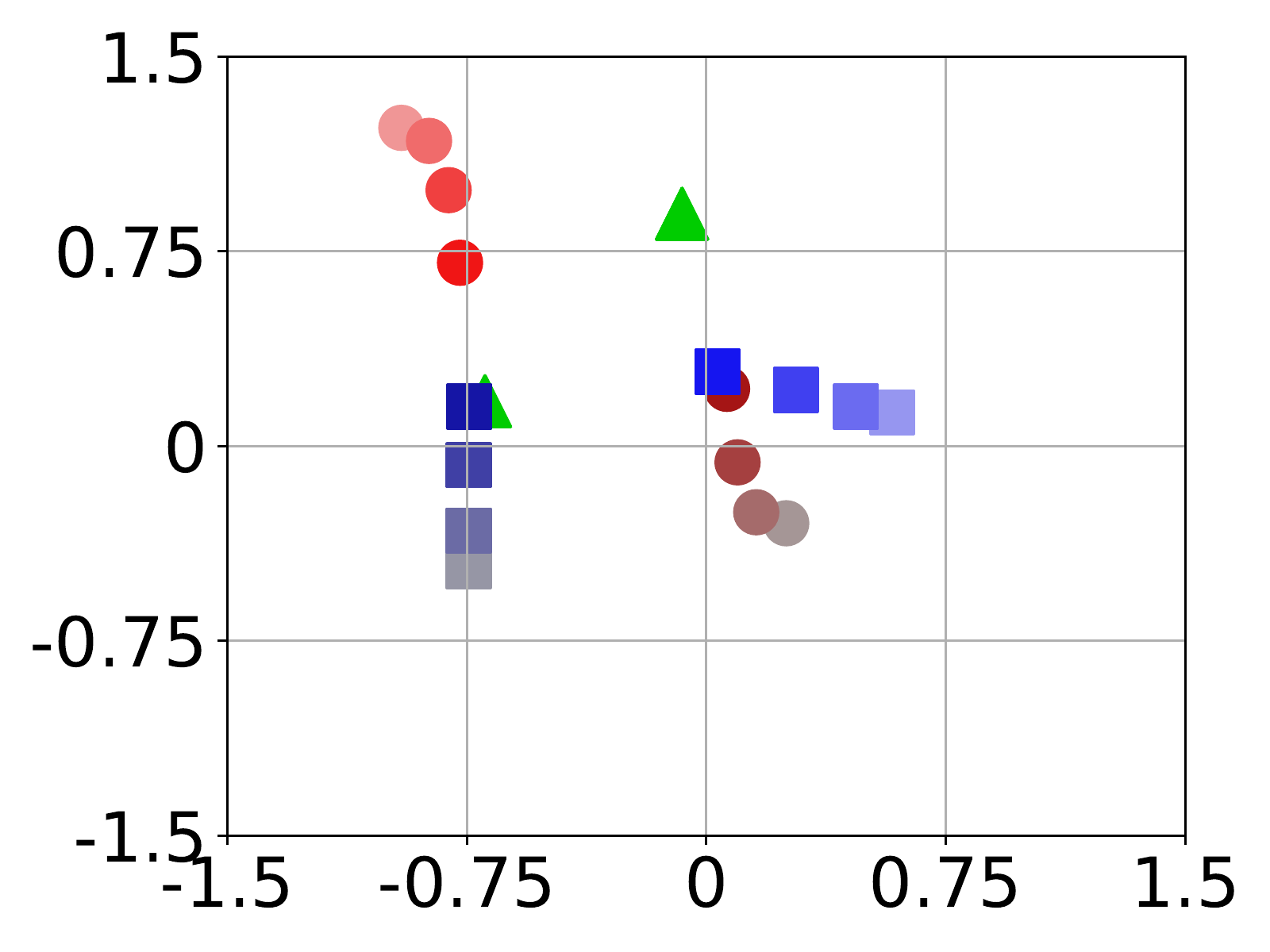}\label{fig:demo130}}\\	
	\vspace{-5mm}
	\subfloat{\includegraphics[width=.4\textwidth]{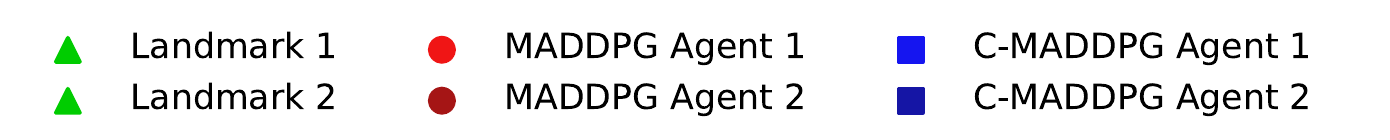}}
	\caption{Role emergence of \our\ agents when  played against \bsln\ agents. The figures have snapshots of various timestamps merged together - agents' positions in the earlier snapshots are in lighter shade. We observe  both \bsln\ agents are interested in moving towards the landmark (\cref{fig:demo129}), whereas \our\ agents wisely split the roles of go-for-landmark and stop-the-opponent between the team members (\cref{fig:demo130}). 
	}
	\label{fig:demo}
\end{figure}

\begin{figure*}[!h]
	\centering
	\subfloat[Reward]{\includegraphics[width=0.17\textwidth]{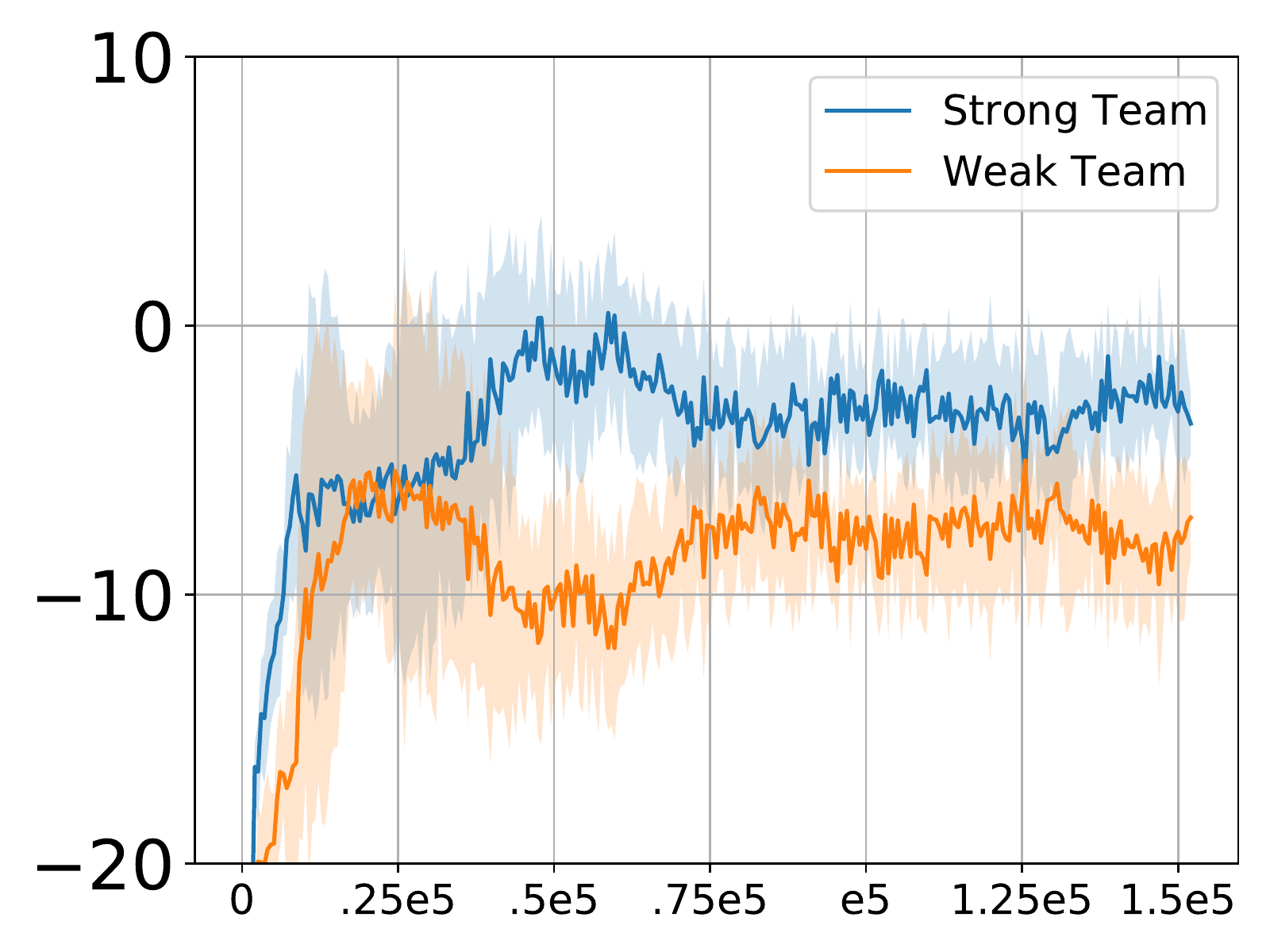}\label{fig:team_wise-score}}\hspace{1cm}
	\subfloat[Landmark Count]{\includegraphics[width=0.17\textwidth]{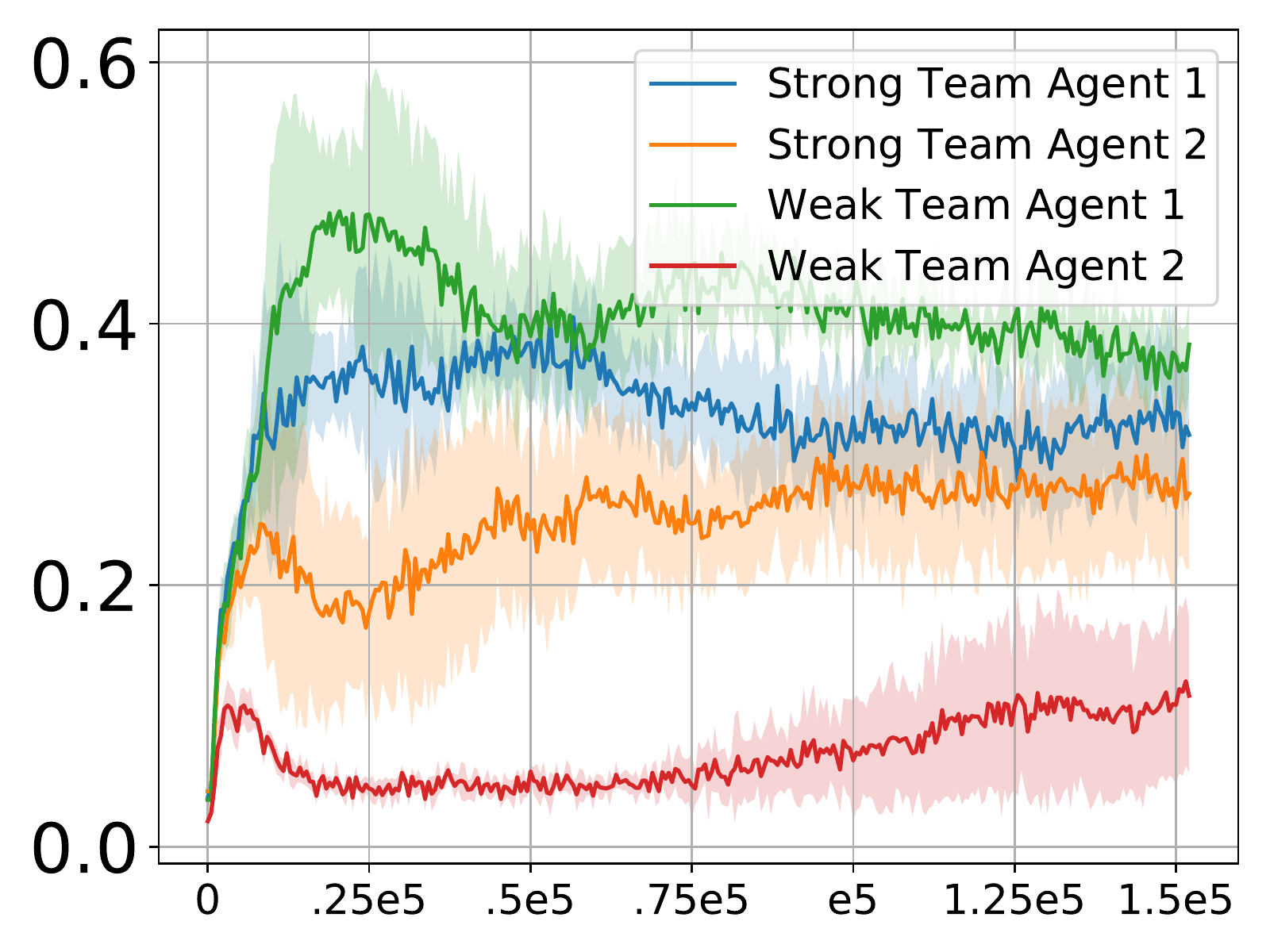}\label{fig:team_wise-count}}\hspace{1cm}
	\subfloat[Win Policy Usage ]{\includegraphics[width=.17\textwidth]{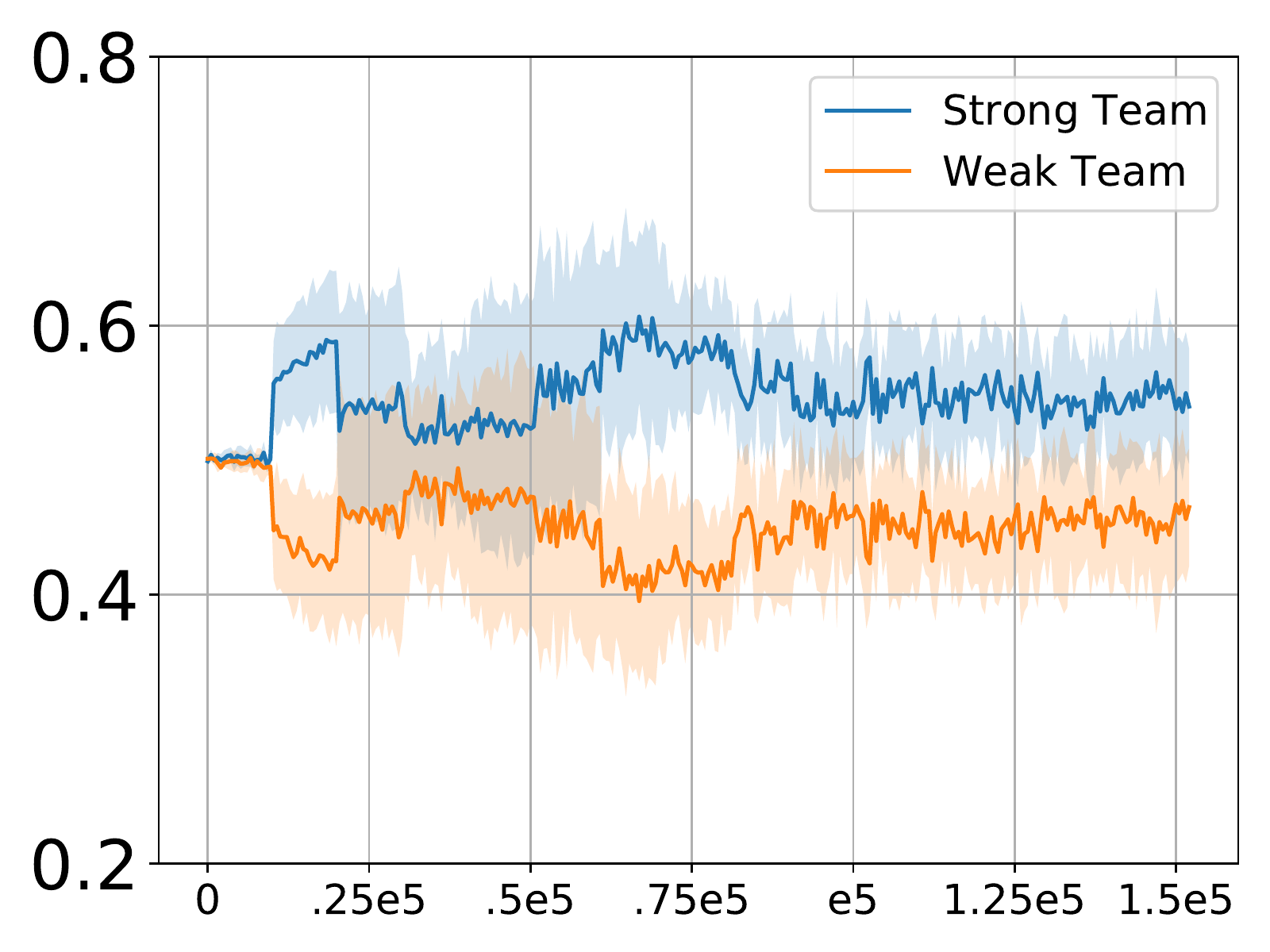}\label{fig:team_wise-win}}\hspace{1cm}
	\subfloat[Speed]{\includegraphics[width=.17\textwidth]{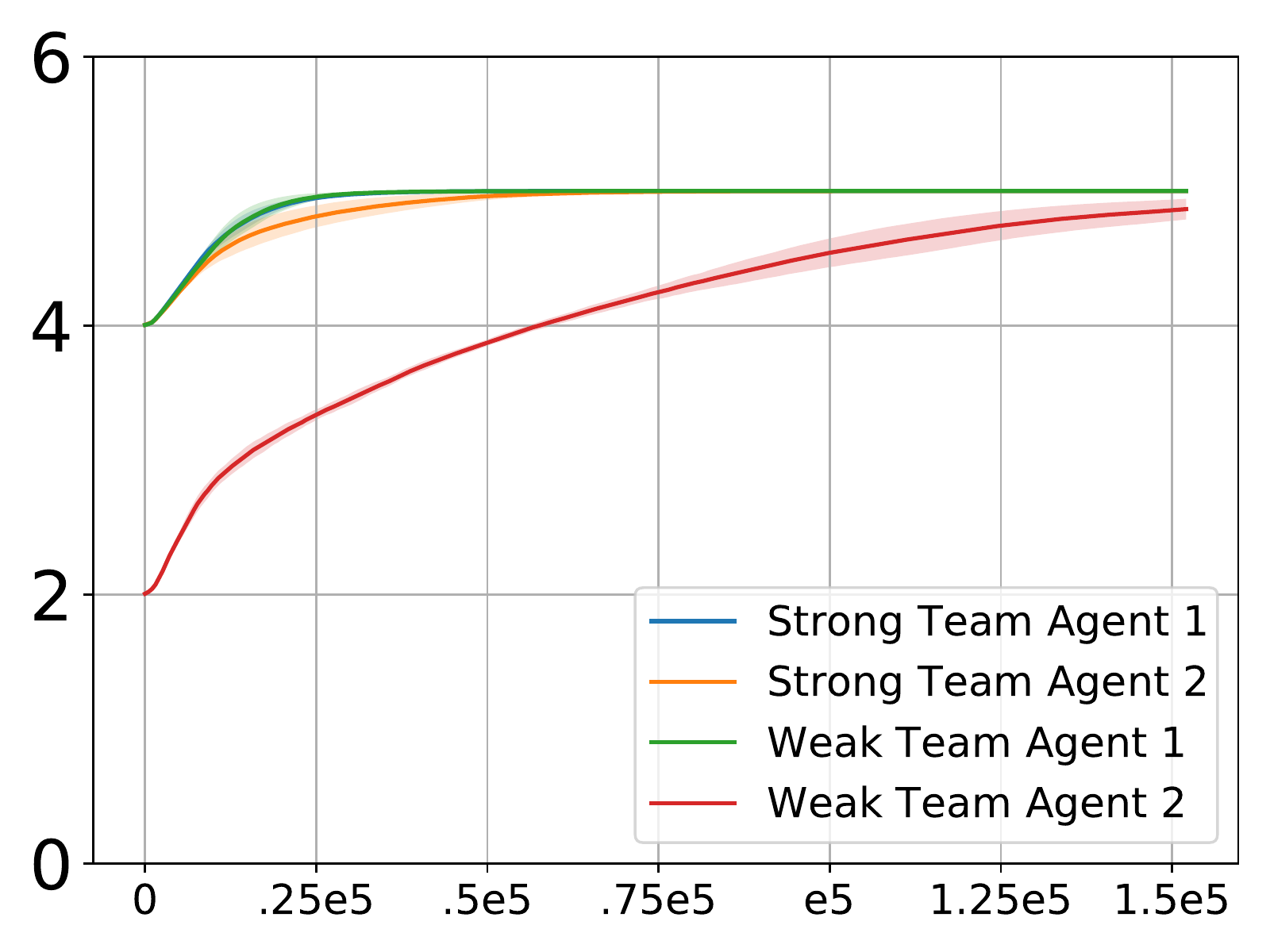}\label{fig:team_wise-speed}}
	
	\subfloat[Reward]{\includegraphics[width=0.17\textwidth]{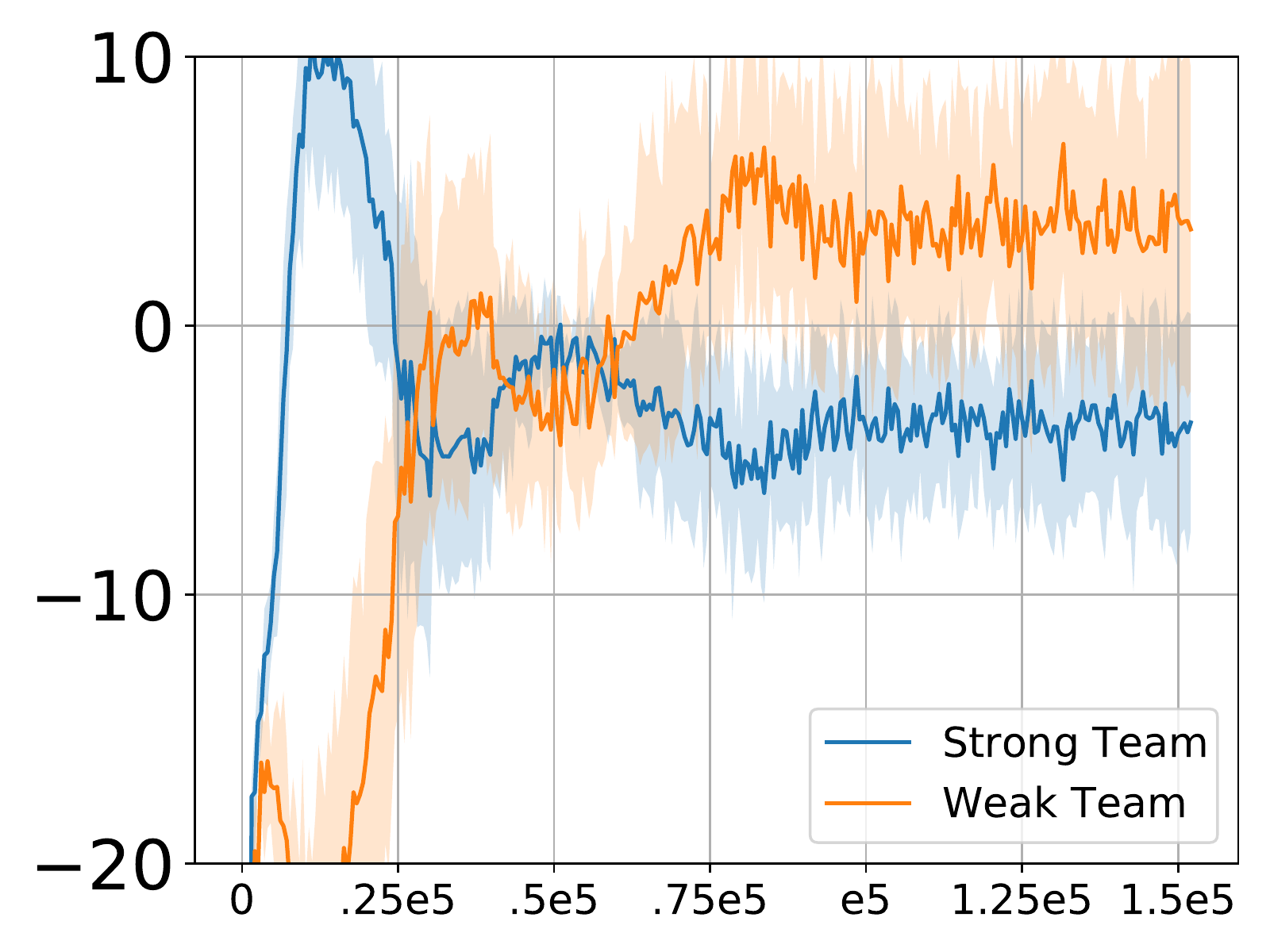}\label{fig:agent_wise-score}}\hspace{1cm}
	\subfloat[Landmark Count]{\includegraphics[width=0.17\textwidth]{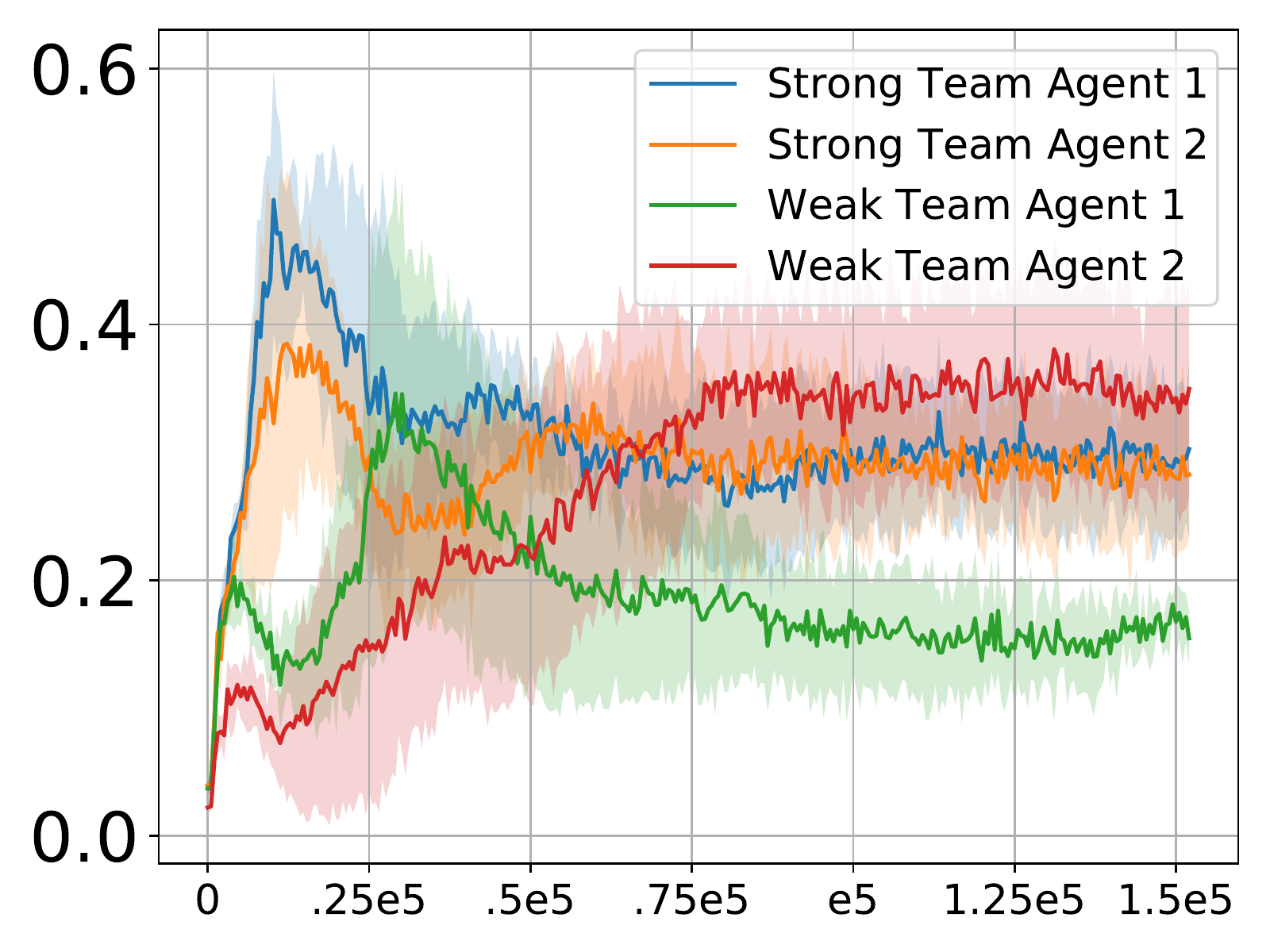}\label{fig:agent_wise-count}}\hspace{1cm}
	\subfloat[Win Policy Usage ]{\includegraphics[width=.17\textwidth]{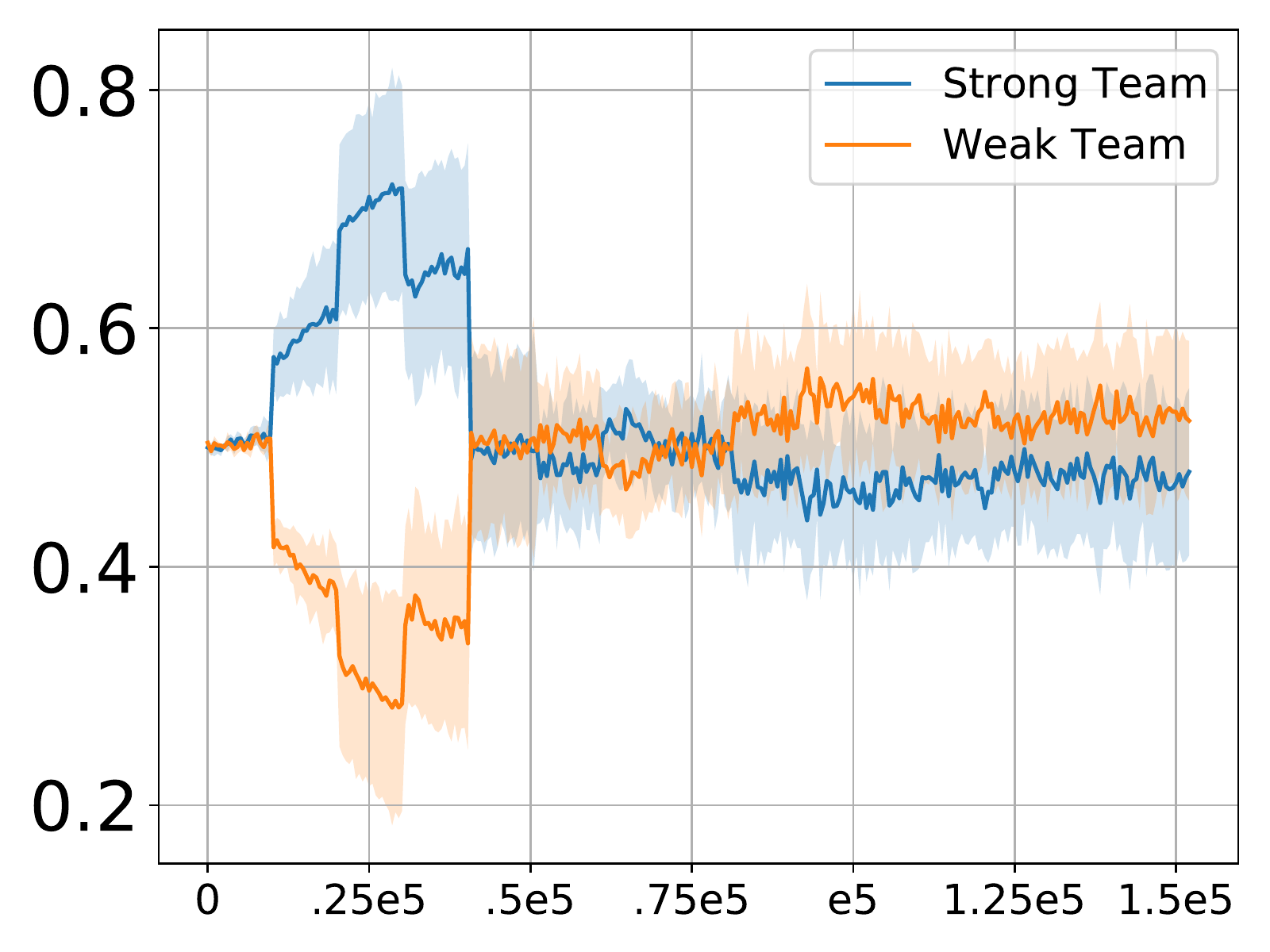}\label{fig:agent_wise-win}}\hspace{1cm}
	\subfloat[Speed]{\includegraphics[width=.17\textwidth]{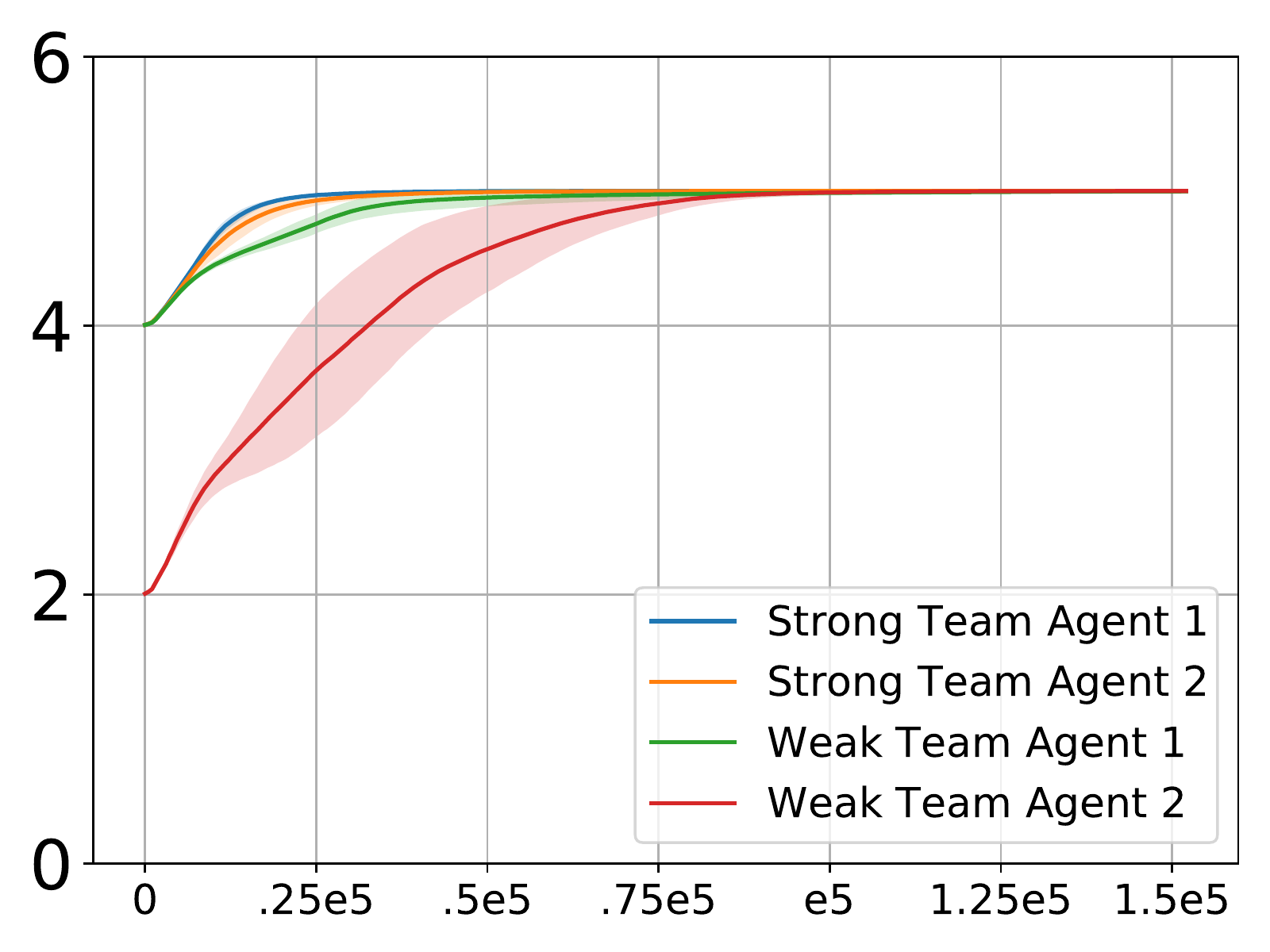}\label{fig:agent_wise-speed}}
	\caption{ Agents trained in \ourgame\  game using \our\ [Fig (a)-(d) for Team-wise Incentive scheme  ($\alpha_{\tT} = 0.1$) and fig. (e)-(h) for Agent-wise Incentive scheme ($\alpha_{\tT} = 0.3$ and $\alpha_{\aA} = 0.7 $)]. Average performance over $4$ different seeds has been reported (the shades signify the  $95\%$ confidence region).}
	
	\label{fig:team_agent_wise}
\end{figure*}

\noindent \textbf{Experimental setup}
For designing the classifier module in \our, a multi-layer perceptron has been used, which consists of two hidden layers with $64$ and $32$ neurons respectively, followed by a single node at the output layer and takes an input of dimension $20$. The hidden layers use ReLU activation function, whereas the output layer uses sigmoid as the activation function. 
For all the experiments here (and henceforth), 
the  landmark touching reward at the end of episode is taken as $r_l=30$, and per timestep penalty for agent $i$, $r_d(i) =  dist\_landmark(i)$ (distance from nearest landmark), the agents move in continuous space, the parameters \texttt{max\_speed} are set to $4$ for all agents, \texttt{MAX\_SPEED} is set to $5$. We train the RL agents for $120000$ episodes for $6$ different seeds and report average results with $95\%$ confidence interval. We have chosen a $3 \times 3$ board. 
	
\noindent \textbf{Performance Comparison of  \our\ and \bsln{:}  }
The comparison between the two algorithms is reported in \cref{fig:calibration-score,fig:calibration-count}. 
While training, after every $1500$ training episodes, agents from both algorithms are made to play among themselves for $1000$ test episodes. 
To remove any bias, 500 initial configurations are generated and two episodes are played with same configuration but  the initial positions of the teams are exchanged.  
It can be seen that as the training progresses, \our\ consistently achieves higher reward with respect to  \bsln\ (\cref{fig:calibration-score}). The landmark reaching statistics also indicates superiority of \our\ team over \bsln\ team on \ourgame (\cref{fig:calibration-count}). 

In order to understand the reason behind  performance difference, we look into the behavior of the agents time-step wise, we find that 
interestingly the two agents of \our\ assume different {\em roles} to ensure that their  team wins, which is not true 
in the case of \bsln{.}
We illustrate this through the \cref{fig:demo}, which presents different snapshots of an episode (merged together) to illustrate the dynamics of the agents. 
\Cref{fig:demo129,fig:demo130} demonstrate two episodes,  where  the initial positions of \our\ and \bsln\ agents are exchanged respectively. Here in \cref{fig:demo130}, one agent  of the \our\ team goes for landmark and another \our\ agent stops an \bsln\ agent from reaching another landmark;  
in \cref{fig:demo129} \bsln\ agents, in a similar situation, do not show any particular trend of role separation. 
In order to understand the importance of collision, we consider all those episodes where
\our\ team wins in both a configuration as well as its reverse and we find, in those cases, the collision rate  is $0.512$, significantly higher than the average 
rate of $0.257$.

Summarizing, it is observed that \our\ agents wisely split the emerged roles, \textit{go-for-landmark} and \textit{stop-the-opponent},  between themselves
resulting in superior performance as seen  in \cref{fig:calibration-score,fig:calibration-count}. 
The reason behind the ability of \our\ in role splitting while adopting winning policy, may be attributed to the design 
of the classifier which leads to the division of sample space between the competing policies. 
The subdivision allows more focused exploration whereby agents  closer to landmark try to reach the target while it is advantageous 
for the team if the other agent  tries to stop the opponent. 

The role emergence is an important requirement for the efficient functioning of a team;  however, in various cases, 
some roles may emerge as the main tasks and the others get relegated to auxiliary services. 
This is true for \ourgame\ where \textit{go-for-landmark} which results in touching the landmark is more 
important as  the speed level increases only after touching the landmark, thus \textit{stop-the-opponent} plays an assistive role. 
So, the strength of an agent can be attributed to the frequency at which it touches a landmark; we will use this knowledge while
understanding and devising fair incentive mechanism in the next section.

%% file: 030Algorithm.tex
\section{Fair Competition}
\label{sec:faircompetition}
In this section, we play \ourgame\ with two teams having unequal \textbf{skill} (represented here by speed)
and try to devise an incentive scheme (catered towards the weak team) to match the final reward of the two teams. In the stronger team, we set the initial \texttt{max\_speed$_i$} of both members to $x$. 
In the weaker team, the speed ({\texttt{max\_speed$_i$} and speed are used interchangeably})
 of one member is set to $x$  whereas the speed of the `weaker' member is set to $y$ with $y < x$. 
We choose this mixed setting for the weaker team 
as that would reveal the more interesting case where inequality is there both within and between
 teams.
For all of the following experiments, all the agents learn their policies 
using \our\ for $T = 150000$ episodes. We perform experiments for $4$ different seeds and finally report the average performance over all seeds for each metric along with the interval region signifying $95\%$ confidence. We report the average episode rewards for each team (\textit{Reward}), the average number of times each agent has touched the landmark per episode (\textit{landmark-count}), the fraction of times a team has used the winning policy (\textit{win-policy-usage}) and instantaneous \texttt{max\_speed$_i$} i.e. \textit{speed} for last $T$ episodes throughout the RL learning episodes~(These metrics are reported 
through [\cref{fig:team_agent_wise} - \cref{fig:intrinsic-rl}]). Table \ref{tab:notations} contains list of important notations used in defining consequtive incentive schemes.
We first propose an incentive scheme targeted towards the weak team.

\begin{table}
	\begin{tabular}{c l}
		\toprule
		Symbol & Meaning \\\midrule
		$\alpha_{\tT}$ & Fraction of additional reward if weak team touches\\
		&   landmark \\
		$\alpha_{\aA}$ & Fraction of additional reward if weak agent touches\\
		&  landmark \\
		$n_{\tT(x)}$ & Performance/speed of specific team $x$(strong/weak)\\
		$n_{\aA(x,\tT(y))}$ & Performance/speed of specific agent $x$(strong/weak) \\
		& of team $y$(strong/weak)\\
		\bottomrule
		\vspace{1mm}
	\end{tabular}
	\caption{List of important notations}
	\label{tab:notations}
\end{table}

\noindent\textbf{Team-wise Incentive :} 
In the changed reward scheme with incentive, every time the strong team touches a target 
(event also called \textit{goal}), it gets the usual reward $r_l$; while for every goal, the weak team 
gets a reward of  $(\alpha_{\tT} + 1 ) \cdot r_l$  where $\alpha_{\tT} \ge 0$.	
All other rewards and penalties remain the same.
The parameter  $\alpha_{\tT}$ can be set manually through trial and error, by best balancing the  cumulative rewards (of the last $T$ instances) of the two teams.

In  [\cref{fig:team_wise-score} - \cref{fig:team_wise-speed}] we have summarized a representative observation at $\alpha_{\tT}=0.1$.
\Cref{fig:team_wise-score}  shows that for the  specific value of $\alpha_{\tT}$ the cumulative rewards can be balanced only for a short duration;  eventually the \textit{stronger team takes over and consistently outperforms the weaker team}.
\Cref{fig:team_wise-count} reveals that the weakest player is reaching the landmark very rarely compared to strong players. Hence it hardly learns how to reach landmark and 
 win episodes.
Consequently, the players of the weaker team progressively start choosing the winning policy less frequently (\cref{fig:team_wise-win}), and skill-level (speed)  of the weaker player also increases at a far slower rate than other players (\cref{fig:team_wise-speed}).
We see that the strong player of the weak team initially performs well, as good as the strong player of the strong team (it also initially quickly increases its speed (\cref{fig:team_wise-speed})), this is because, within its team, it gets more opportunity.
But without the help of the weak player,
its performance progressively deteriorates against a team where both members are continuously improving. We conclude that in addition to team incentive, {\bf special incentive is needed for the weak agent}  to balance the long-term total reward.

To motivate such incentive in real life, let us take the example of a sales team where members have to perform various roles like maintaining paperwork and doing the actual sales to customers;  the framework marks the paperwork role as less important. Cash incentive is provided to the team when actual sales are performed. The iota of incentive depends upon the team and individual member’s expertise, the less experienced member fetches more reward on performing the actual sales. Of course, the reward is then distributed equally among all the members of the team.  The mechanism nudges the team to let the less experienced do the actual sales
as that would fetch higher reward to the entire team
if successful; in this process, she learns how to perform the job and become trained thus helping the team/herself in the longer run. Without such differentiation, the weaker member would always be made to do the mundane back-office paperwork and never get a chance to learn field operations.

\begin{figure*}[!t]
	\centering
	\subfloat[Reward]{\includegraphics[width=.20\textwidth]{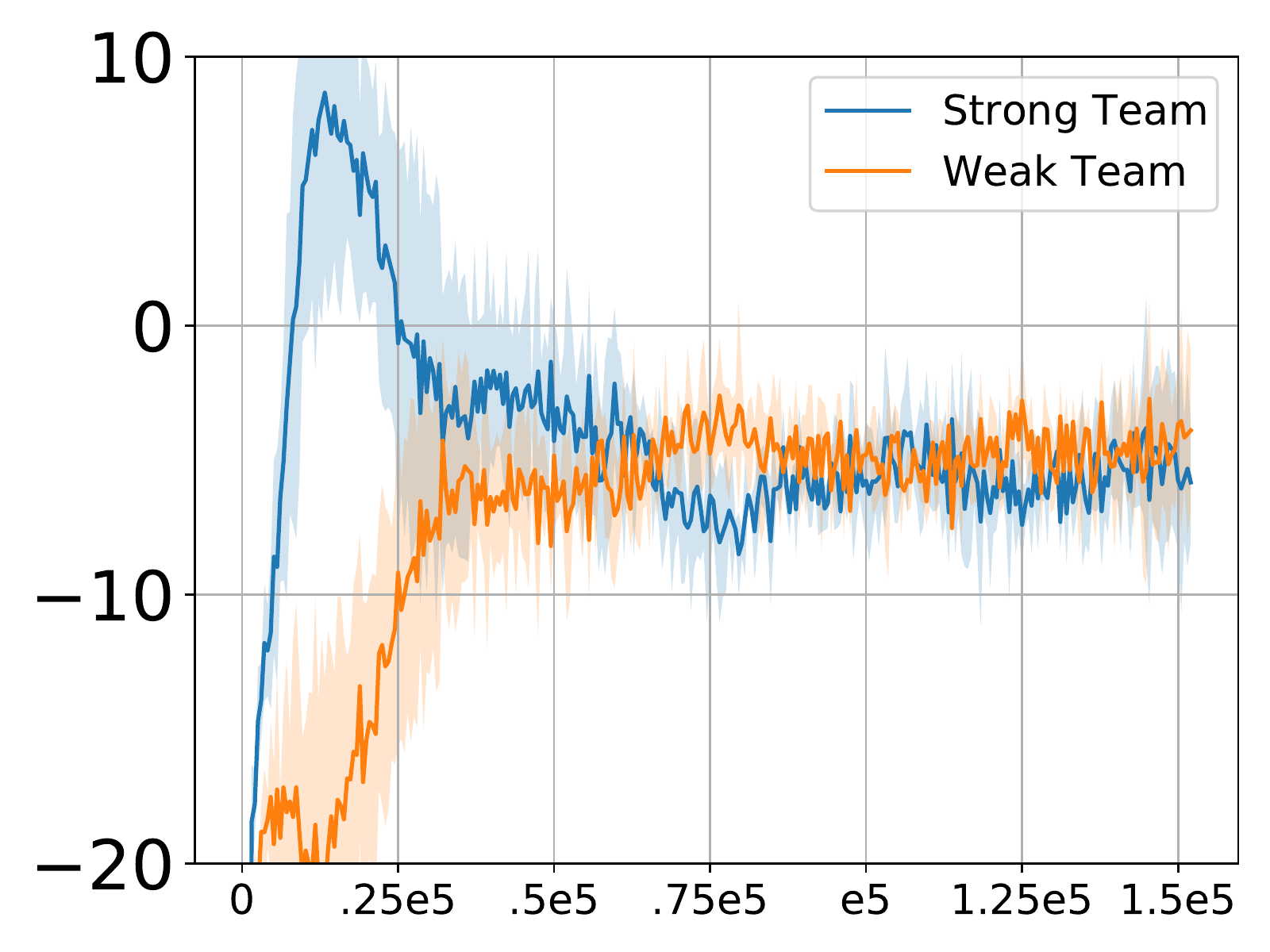}	
		\label{fig:intrinsic-landmark-score}}
	\subfloat[Landmark Count]{\includegraphics[width=.20\textwidth]{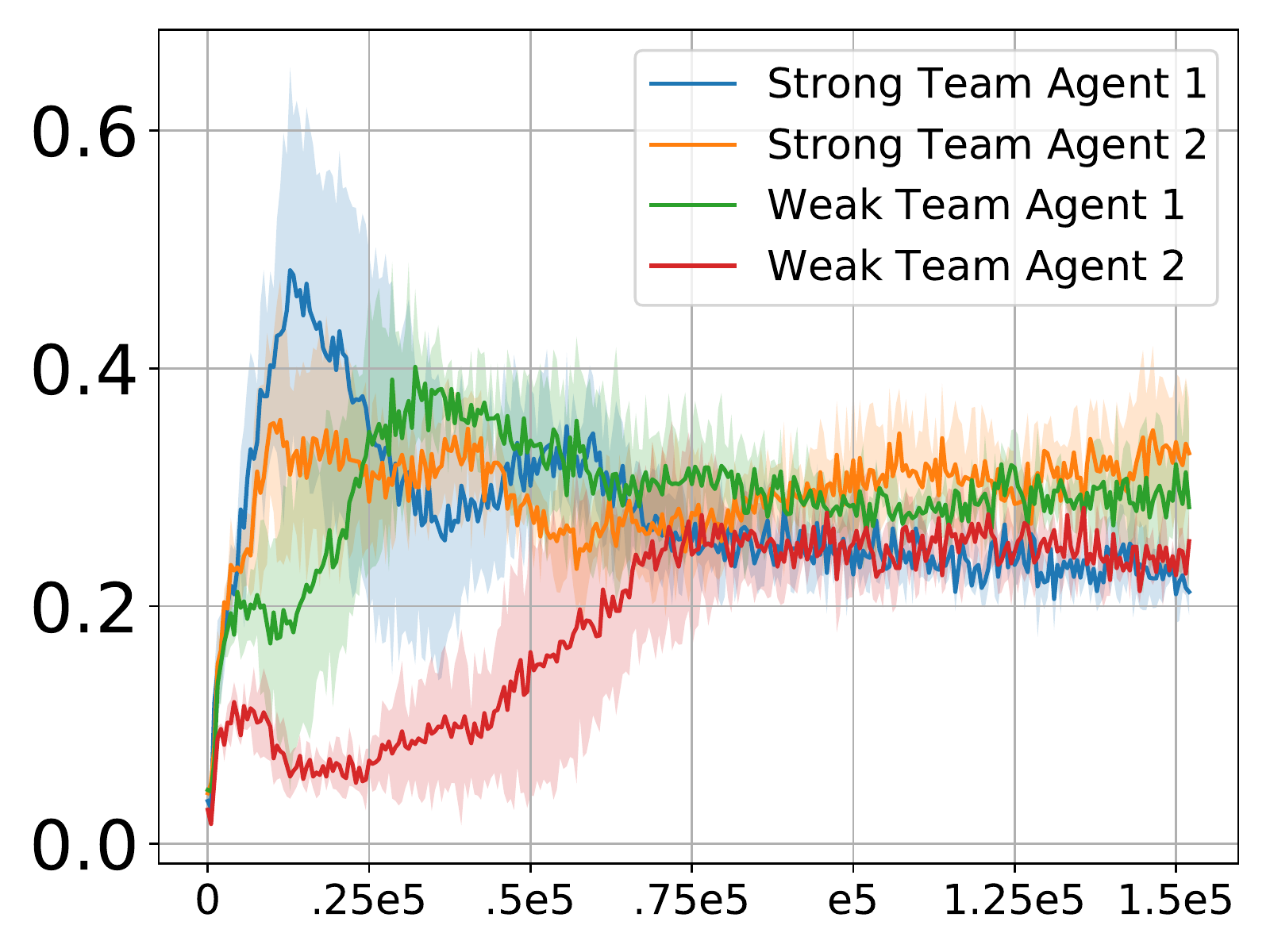}
		\label{fig:intrinsic-landmark-count}}
	\subfloat[Win Policy Usage ]{\includegraphics[width=.20\textwidth]{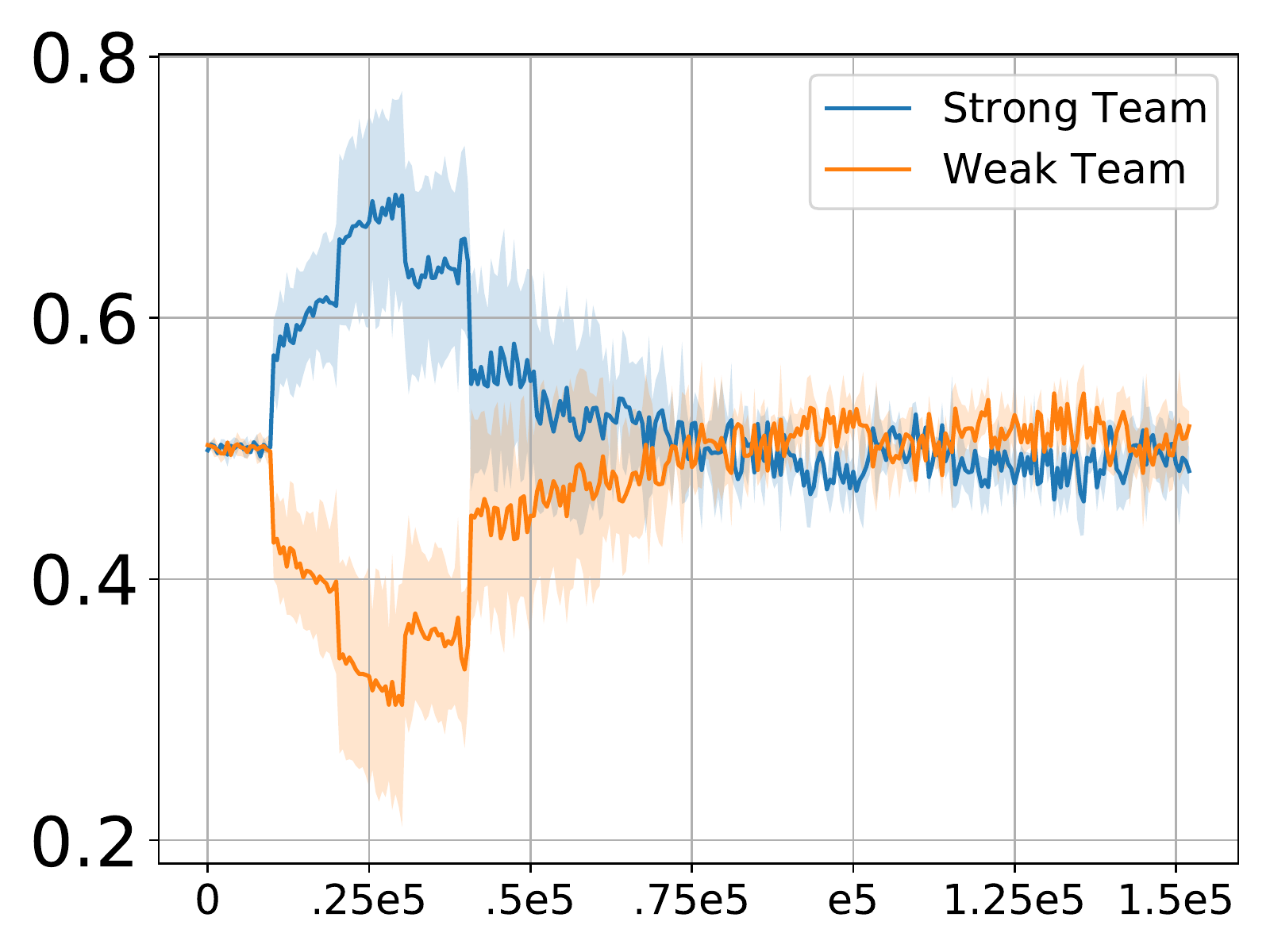}
		\label{fig:intrinsic-landmark-win}}
	\subfloat[Speed]{\includegraphics[width=.20\textwidth]{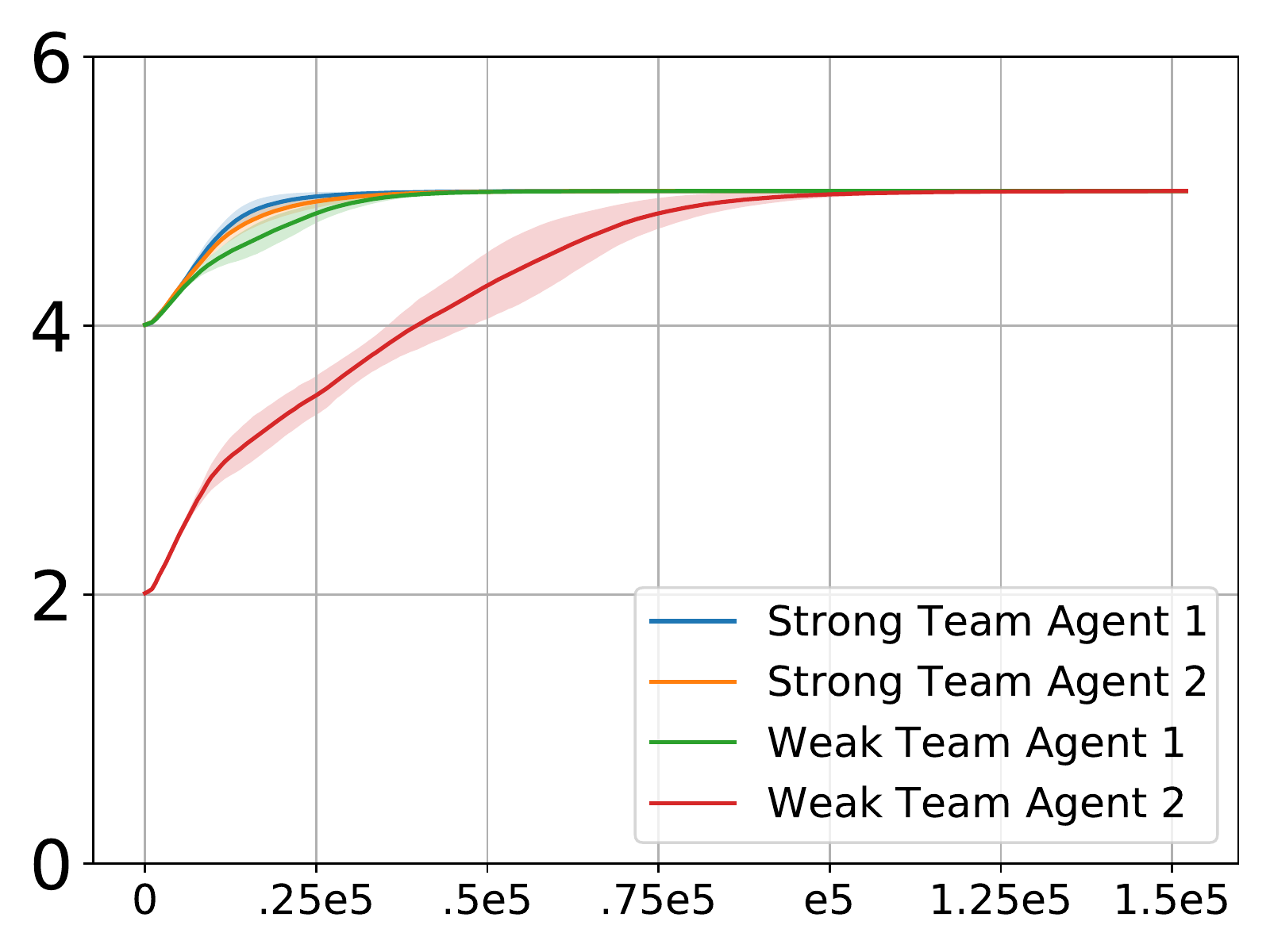}
		\label{fig:intrinsic-landmark-speed}}
	\subfloat[Incentive]{\includegraphics[width=.20\textwidth]{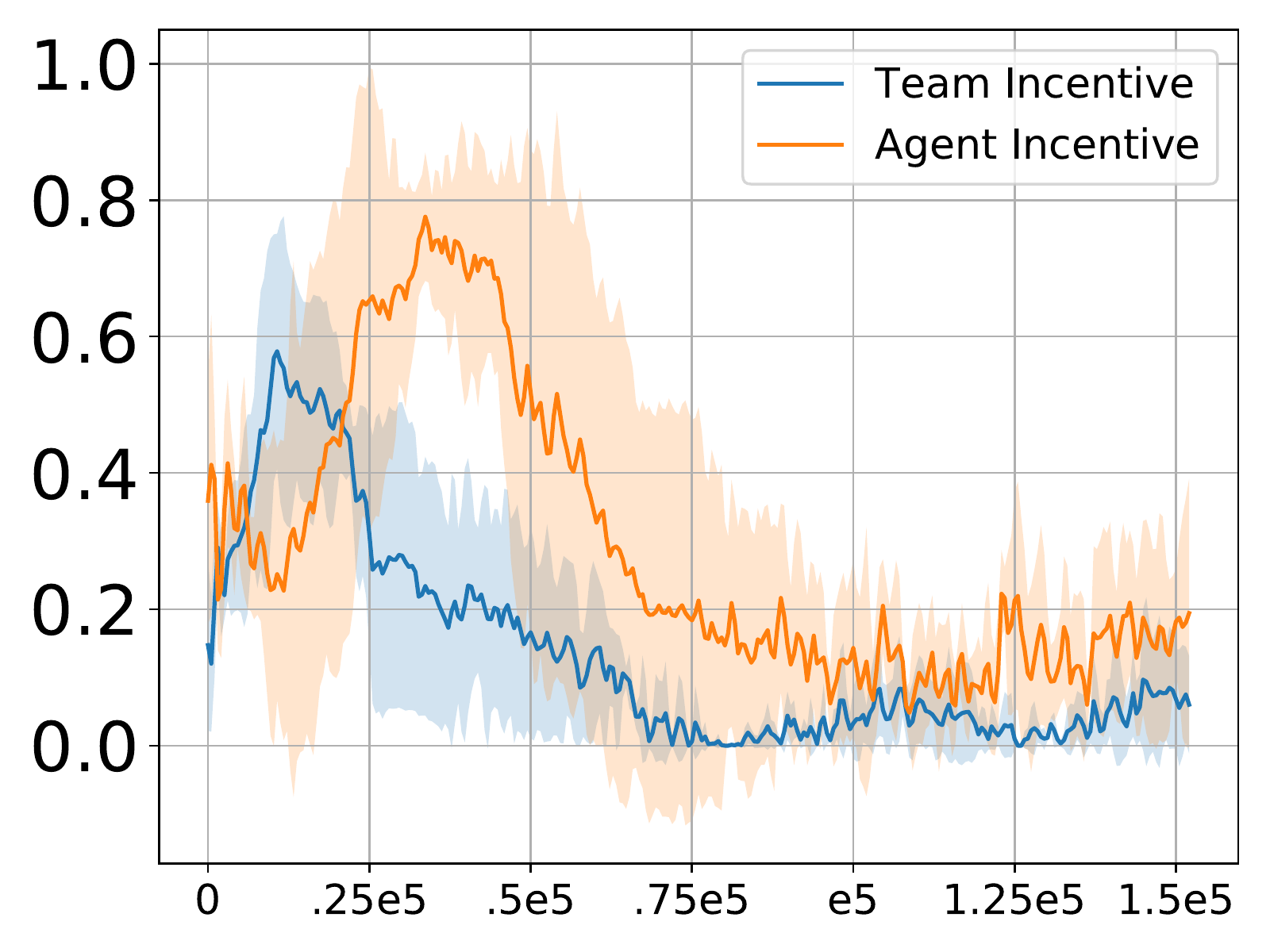}
		\label{fig:intrinsic-landmark-incentive}}
	\\
	\subfloat[Reward]{\includegraphics[width=.20\textwidth]{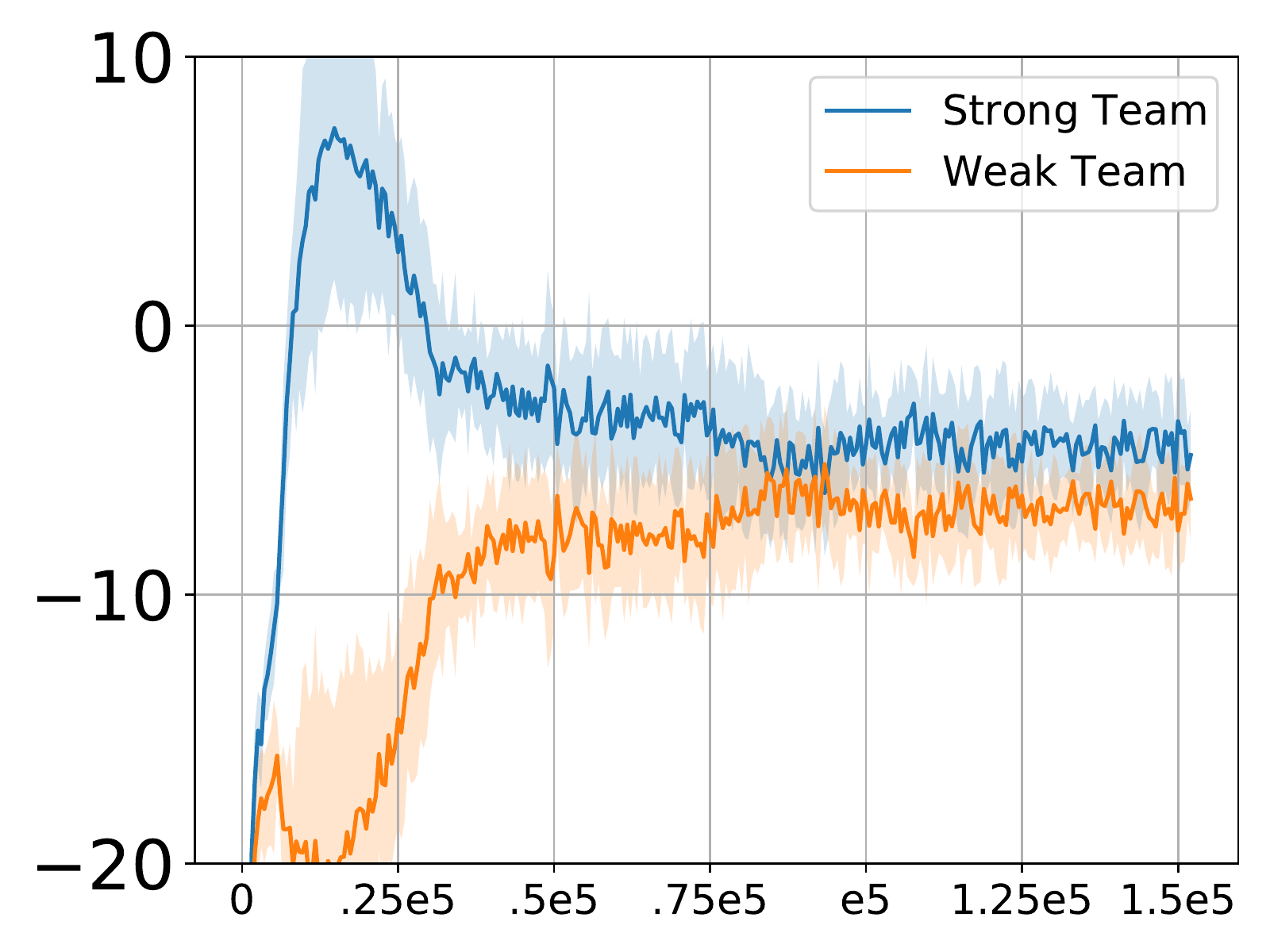}
		\label{fig:intrinsic-speed-score}}	
	\subfloat[Landmark Count]{\includegraphics[width=.20\textwidth]{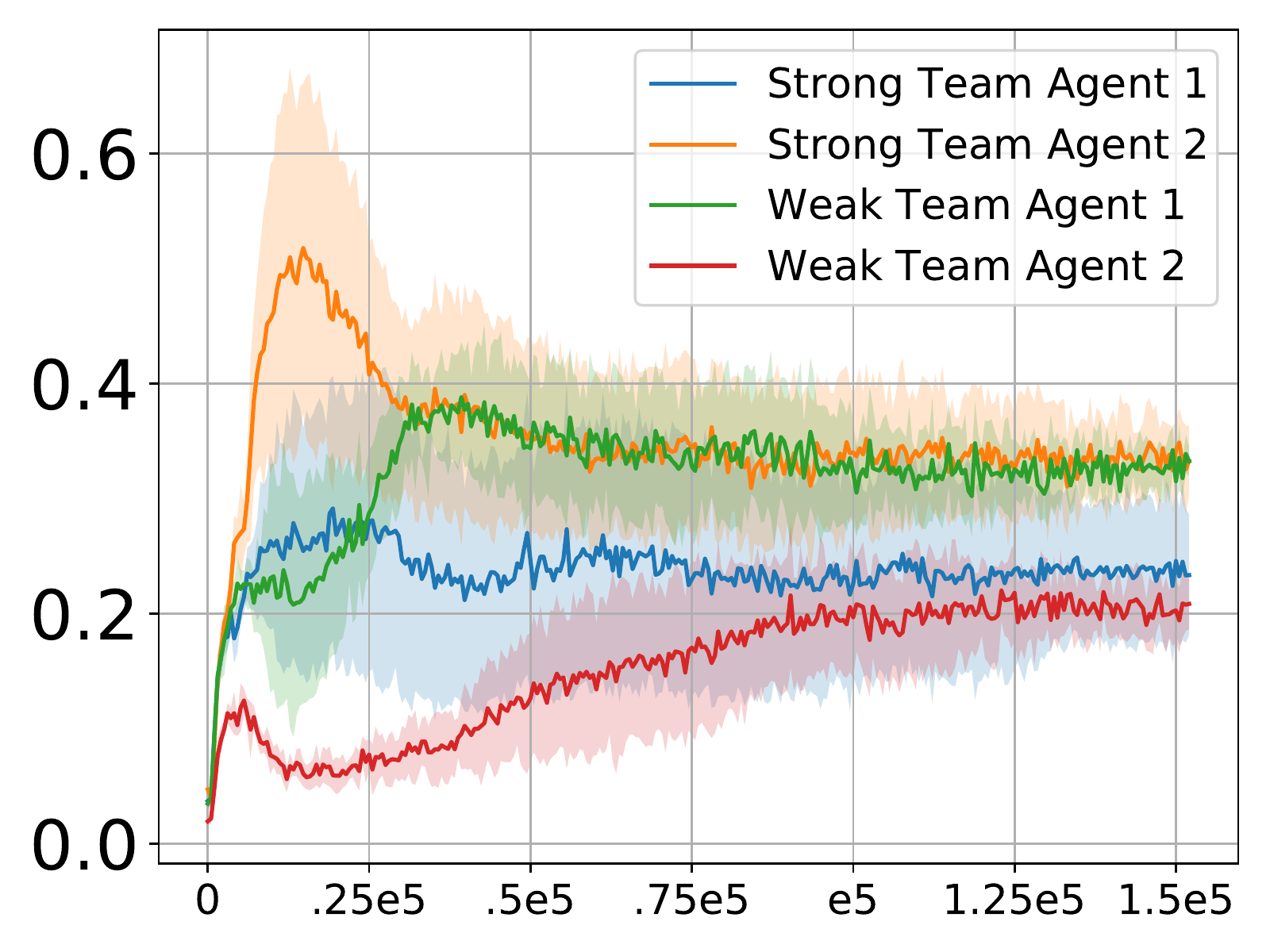}
		\label{fig:intrinsic-speed-count}}
	\subfloat[Win Policy Usage ]{\includegraphics[width=.20\textwidth]{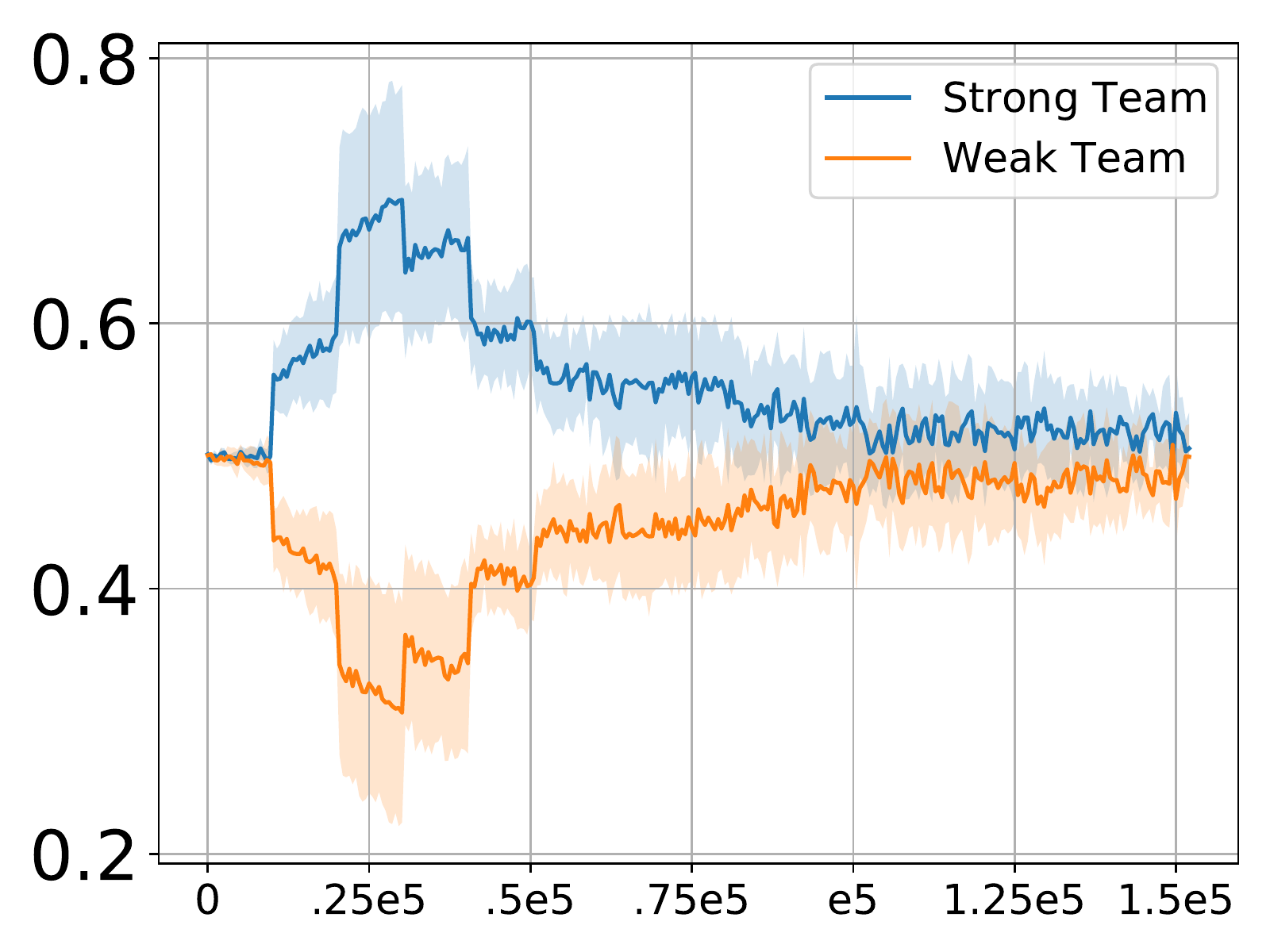}
		\label{fig:intrinsic-speed-win}}
	\subfloat[Speed]{\includegraphics[width=.20\textwidth]{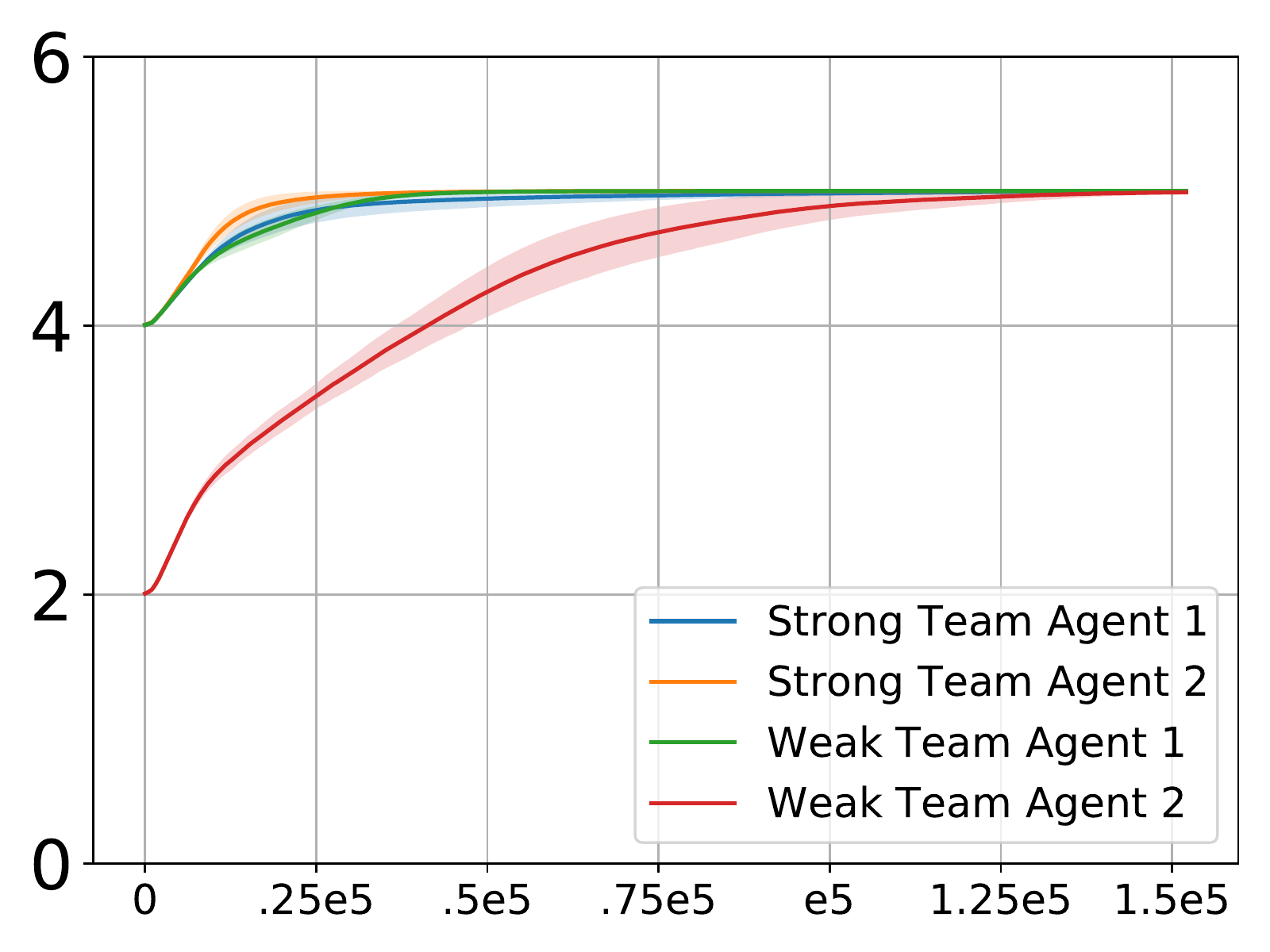}
		\label{fig:intrinsic-speed-speed}}
	\subfloat[Incentive]{\includegraphics[width=.20\textwidth]{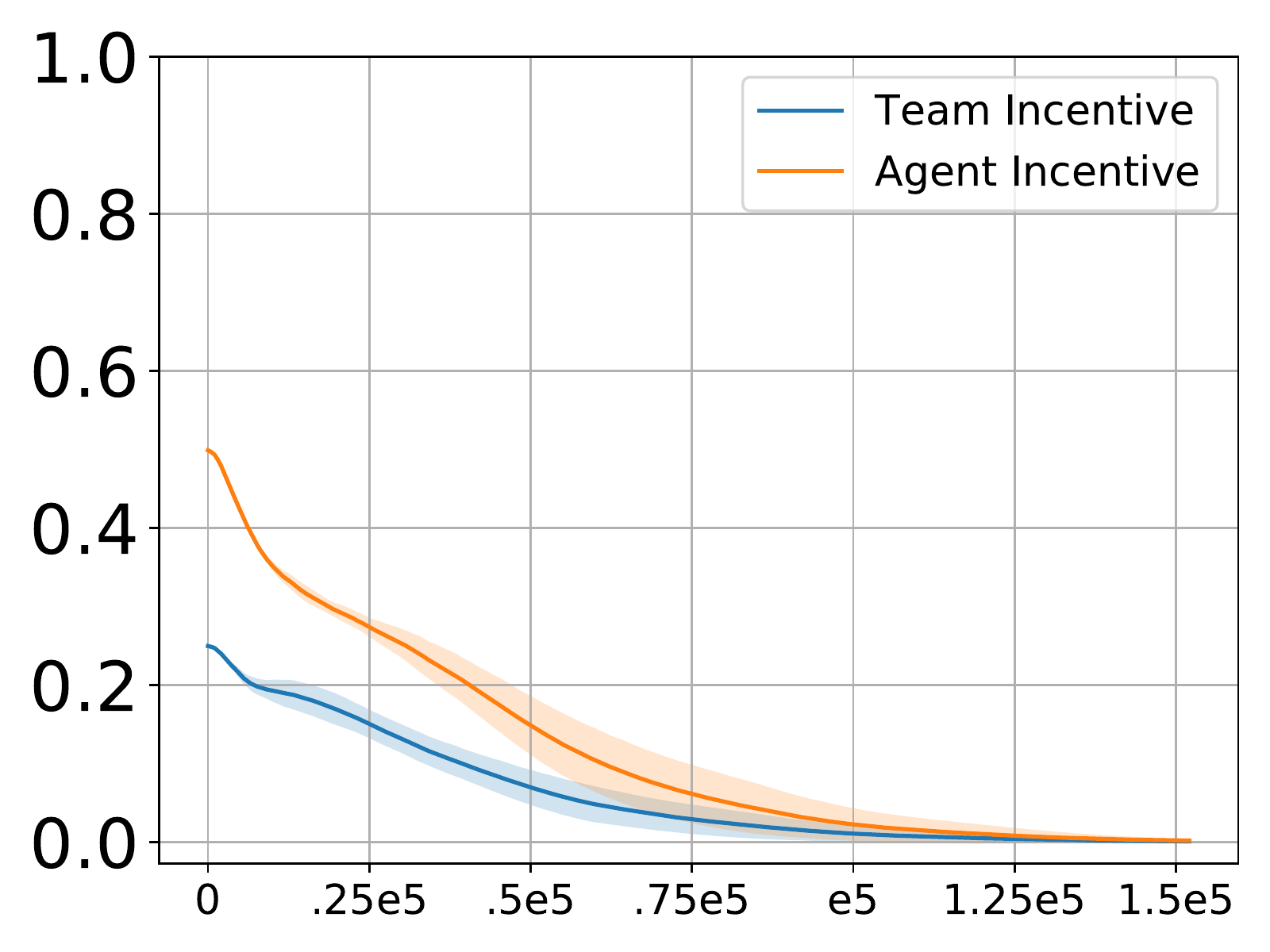}
		\label{fig:intrinsic-speed-incentive}}
	\vspace{-1mm}	
	\caption{Agents trained in \ourgame\ game using \our\  [Fig. (a)-(e) denotes Landmark-based Dynamic Incentive and fig. (f)-(j) denotes Speed-based Dynamic Incentive]. Average performance over $4$ different seeds has been reported (the shades signify the  $95\%$ confidence region).}
	\label{fig:intrinsic-landmark_speed}
\end{figure*}

\noindent \textbf{Agent-wise Incentive:} 
Here we consider a more explicit incentive scheme, namely agent-wise reward. 
In this scheme, the weaker team gets $ (\alpha_{\aA} + \alpha_{\tT} + 1 ) \cdot r_l$ if its weaker member touches the landmark  and  $( \alpha_{\tT} + 1 ) \cdot r_l $ if the stronger member does so; whereas each success of the stronger team is rewarded with $r_l $. 
All other rewards and penalties remain the same. 
The intuition is that such a differential incentive mechanism may help the weak team to discover the huge benefit of the 
weak player touching the target, and consequently the policies would increasingly 
gear the weak player towards attempting to touch the target. 
This in turn would allow the weaker agent to learn and increase her capability (which was not happening in the
previous case) and in the process, the difference between the two teams may disappear. 			

In  [\cref{fig:agent_wise-score} - \cref{fig:agent_wise-speed}] we have summarized a representative observation at $\alpha_{\tT}=0.3$ and $\alpha_{\aA} = 0.7 $. 
Here we observe that in \cref{fig:agent_wise-score}, the {\em weak team}  initially scores less rewards than the strong team, 
but eventually, it {\em gains expertise and outperforms the stronger team in terms of rewards. }
Looking at the corresponding landmark plots, we can conclude that initially mostly the stronger player in the weak team was going for the landmark, but eventually, as the high reward for the weaker agent
is discovered, the weaker agent begins to play a significant role (\cref{fig:agent_wise-count}). Thus we find that unlike 
in the previous case, the winning policy 
is pursued by both the teams equally (\cref{fig:agent_wise-win}). 
The speed of the weaker agent also increases at a much faster rate  than the previous case (\cref{fig:agent_wise-speed}). Thus, in the end, the initial incentive value becomes disproportionate and this results in the continued superior performance of the weaker team. 
This experiment provides a strong evidence that to ensure  sustainable existence and growth, not only the {\bf weak team but also the weak players} need 
targetted  incentive. 

From the study 
we can conclude that an agent incentive successfully improves the weaker team and eventually brings 
it to the level of the stronger team.  So if that incentive is not changed over time, the weaker team would outperform 
the stronger getting `unfair' advantage. Consequently, the incentive needs to be dynamic which would ensure that 
none of the teams get extra benefit at any point in time.  
We will discuss various such dynamic incentive schemes in the next section.

\begin{table}
	\begin{tabular}{c l}
		\toprule
		Name & Incentive scheme \\\midrule
		StaticTeam & $\alpha_{\aA}=0,\alpha_{\tT}=\text{const}$  \\
		\hline
		StaticAgent & $\alpha_{\aA}=\text{const}_1,\alpha_{\tT}=\text{const}_2$ \\
		\hline 
		DynamicLandmark & $\alpha_{\tT} = n_{\tT(\tiny{strong})}-n_{\tT(weak)}$,\\
		&$\alpha_{\aA} = n_{\aA(strong,\tT(weak))}-n_{\aA(weak,\tT(weak))}$\\
		&$n_{(\cdot)}$ measures landmark count.\\
		\hline 
		DynamicSpeed & $\alpha_{\tT} = n_{\tT(\tiny{strong})}-n_{\tT(weak)}$,\\
		& $\alpha_{\aA} = n_{\aA(strong,\tT(weak))}-n_{\aA(weak,\tT(weak))}$\\
		&$n_{(\cdot)}$ measures speed.\\
		\hline 
		Team-RL- & $\alpha_{\tT} =\pi(s)$ \\
		Agent-Dynamic & $\alpha_{\aA} = n_{\aA(strong,\tT(weak))}-n_{\aA(weak,\tT(weak))}$ \\ 
		&$n_{(\cdot)}$ measures speed.\\
		\hline 
		Team-Dynamic& $\alpha_{\tT} = n_{\tT(\tiny{strong})}-n_{\tT(weak)}$ \\
		-Agent-RL &  $\alpha_{\aA}=\pi(s)$\\
		&$n_{(\cdot)}$ measures speed.\\
		\bottomrule
		\vspace{1mm}
	\end{tabular}
	\caption{List of incentive schemes}
	\label{tab:incentive_schemes}
\end{table}

%% file: 050Dynamic.tex
\begin{figure*}[!t]
	\centering
	\subfloat[Reward]{\includegraphics[width=.20\textwidth]{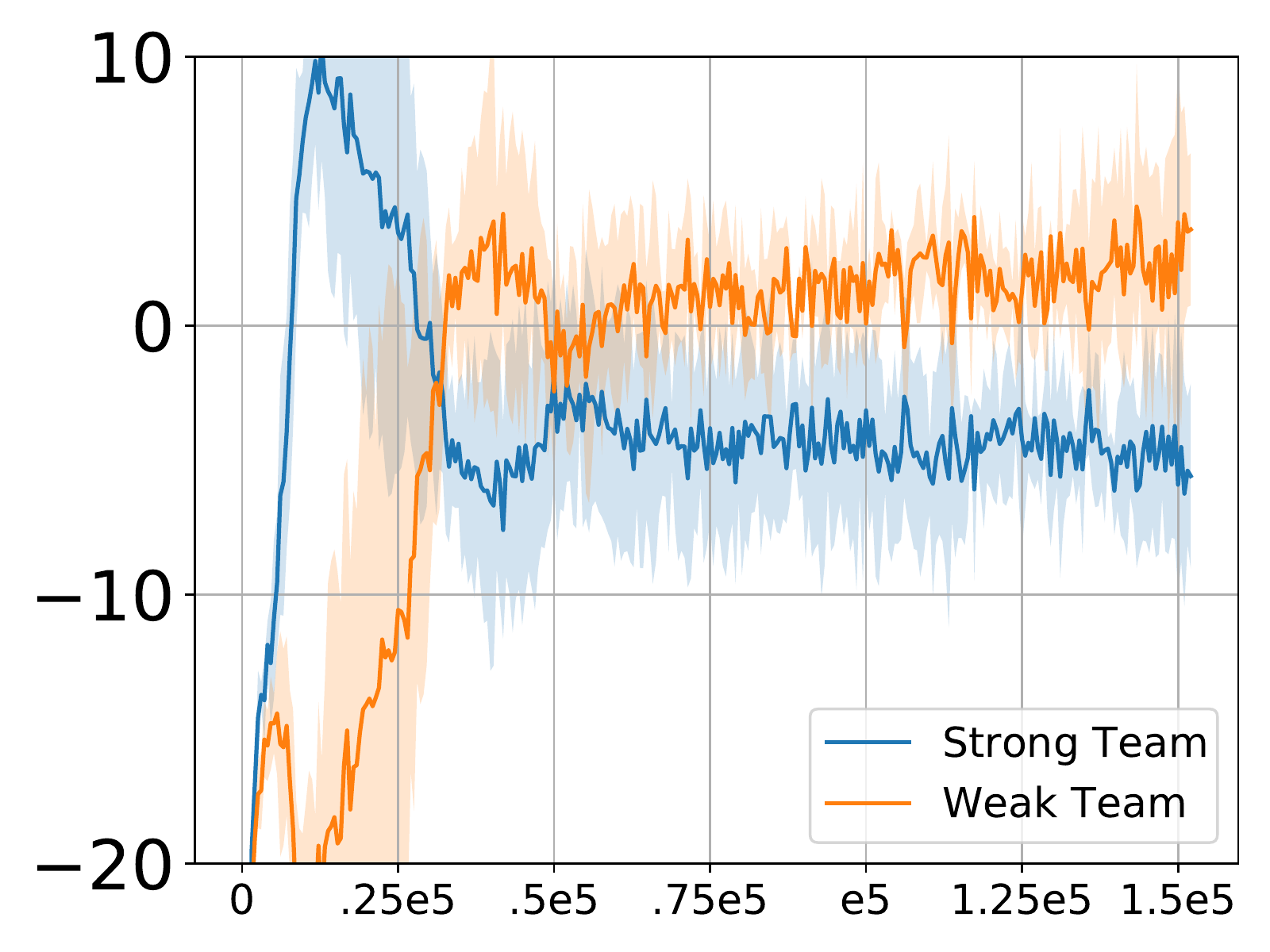}
		\label{fig:team-rl-agent-dyn-w-p-score}}	
	\subfloat[Landmark Count]{\includegraphics[width=.20\textwidth]{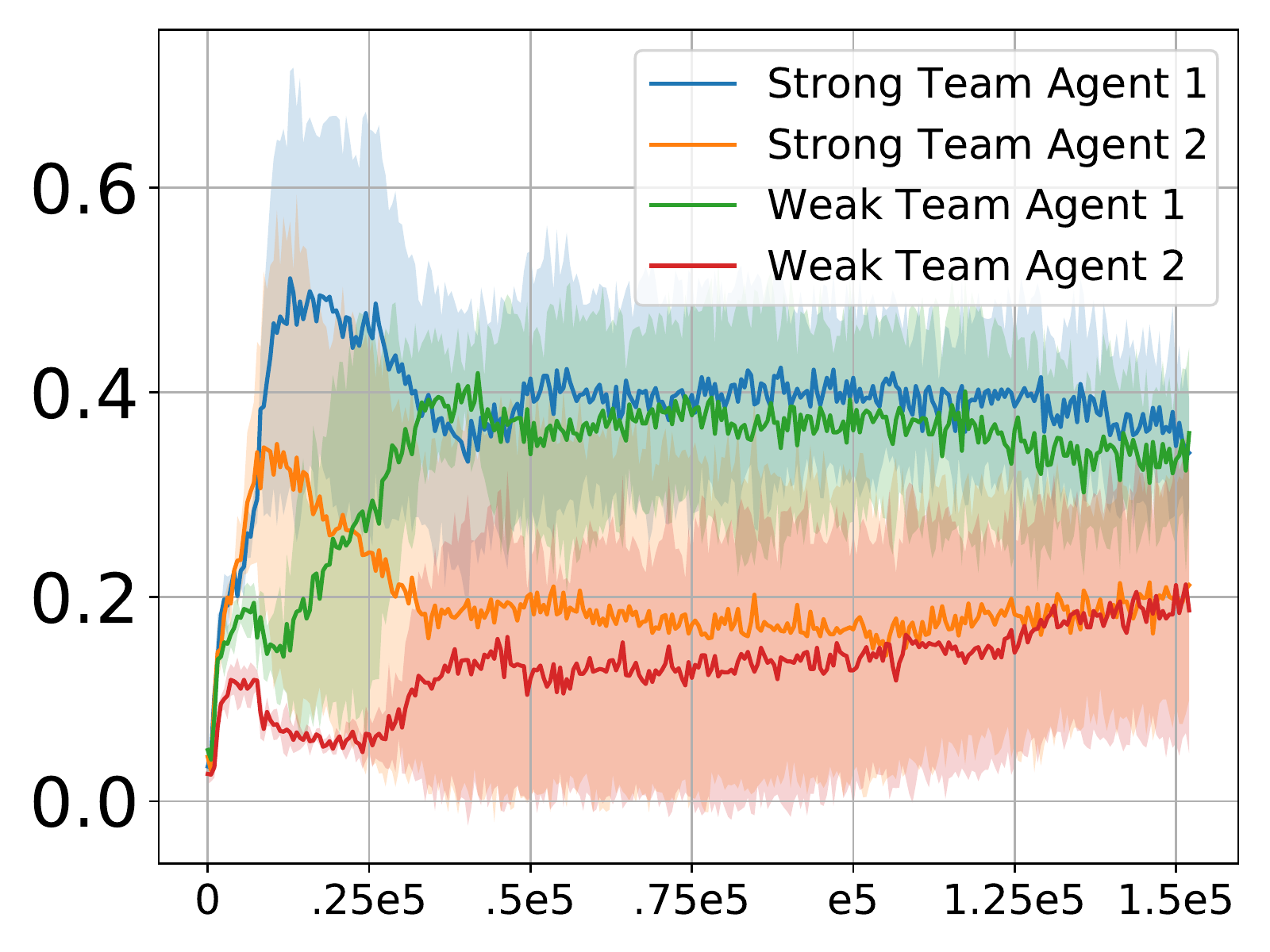}
		\label{fig:team-rl-agent-dyn-w-p-count}}	
	\subfloat[Win Policy Usage ]{\includegraphics[width=.20\textwidth]{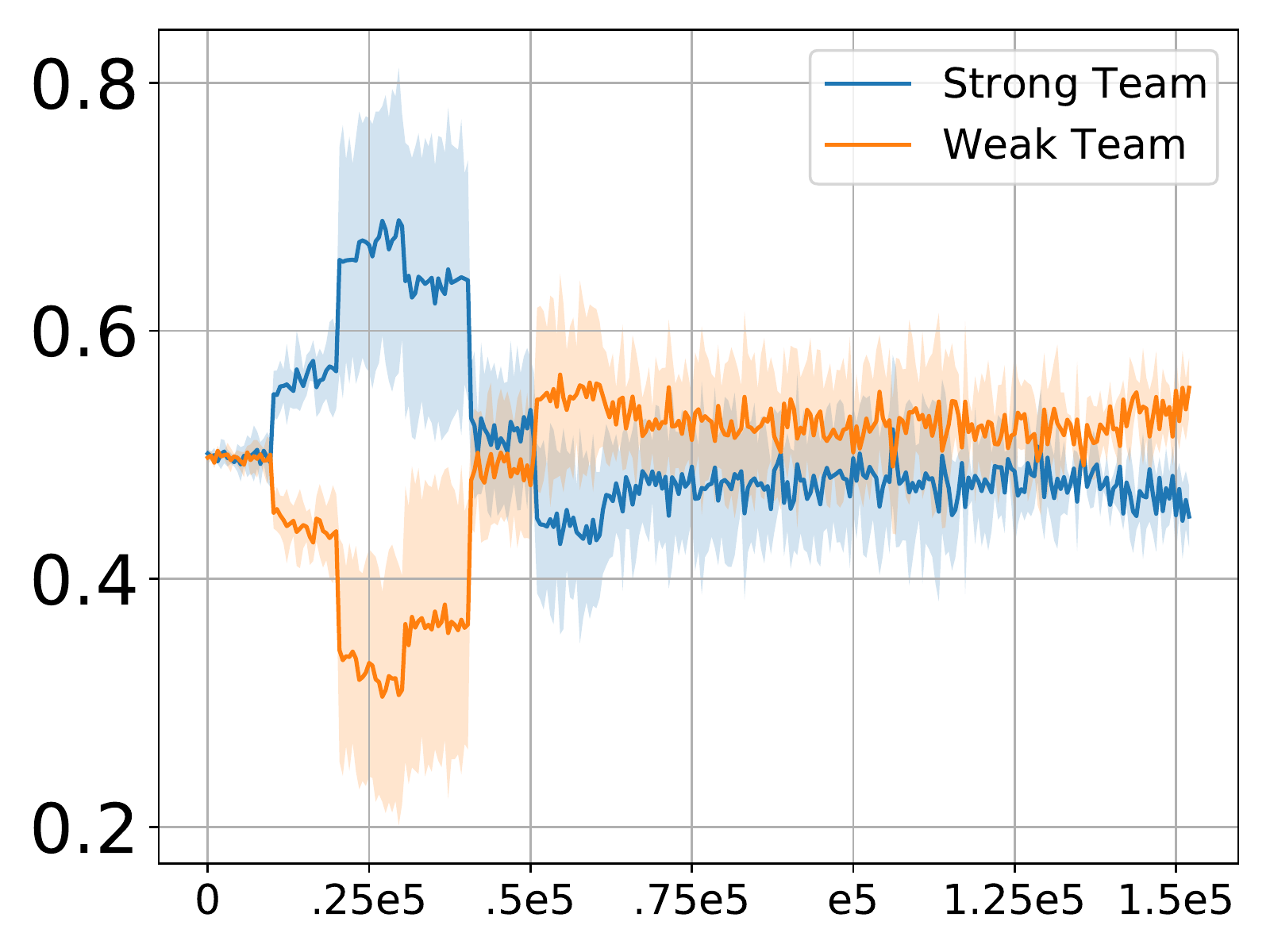}
		\label{fig:team-rl-agent-dyn-w-p-win}}		
	\subfloat[Speed ]{\includegraphics[width=.20\textwidth]{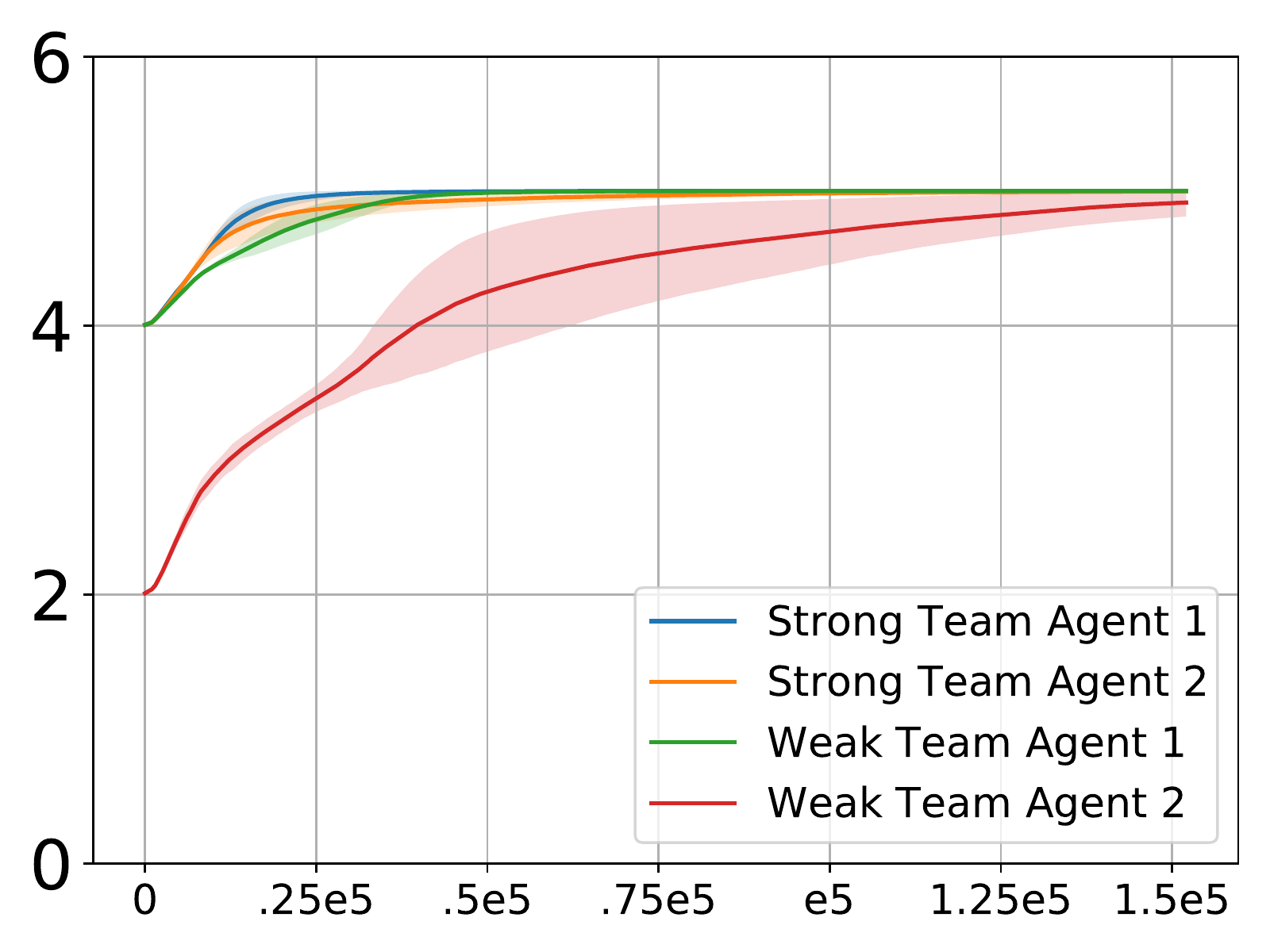}
		\label{fig:team-rl-agent-dyn-w-p-speed}}		
	\subfloat[Incentive ]{\includegraphics[width=.20\textwidth]{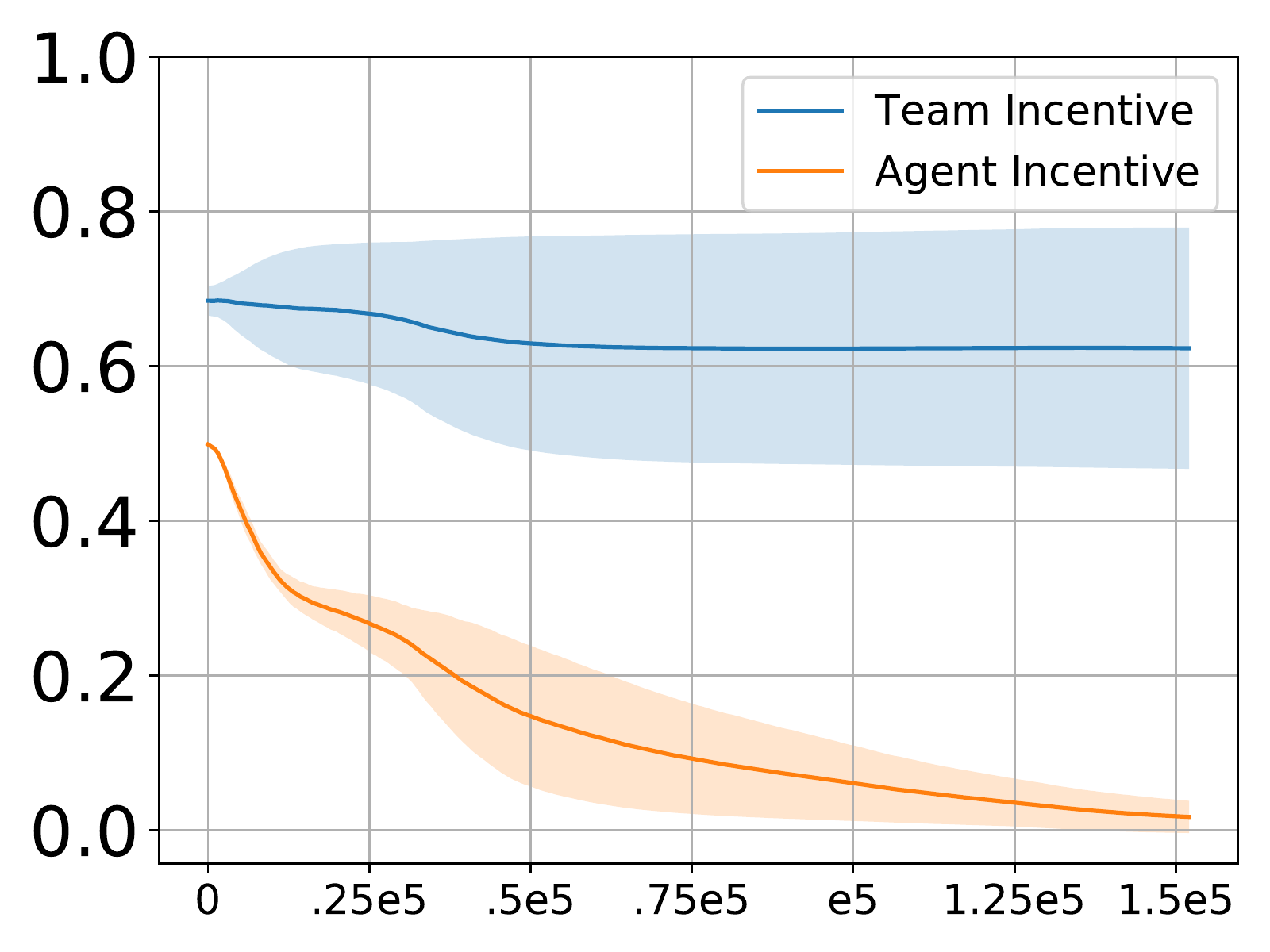}
		\label{fig:team-rl-agent-dyn-w-p-incentive}}		\\
	\subfloat[Reward]{\includegraphics[width=.2\textwidth]{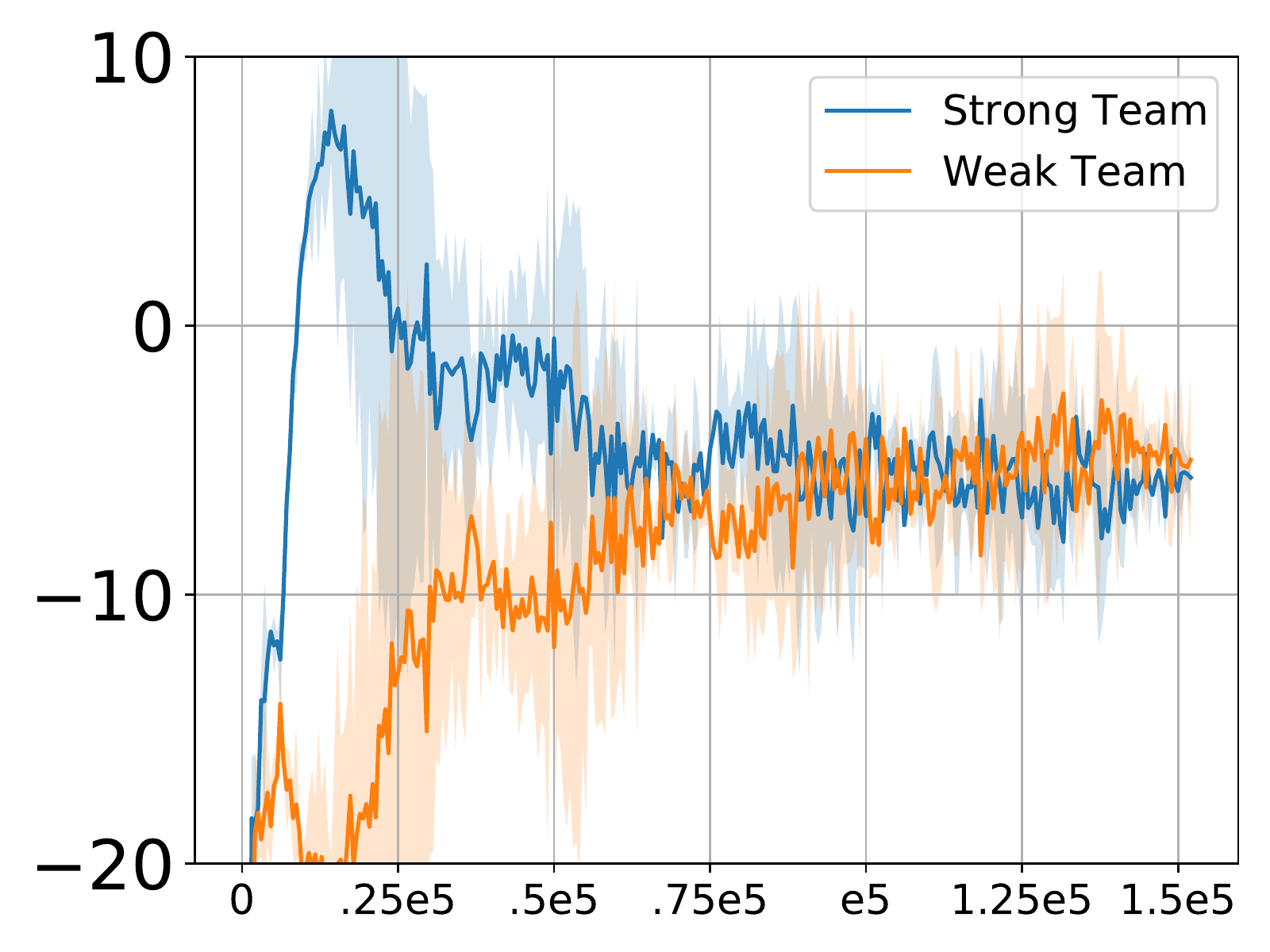}	
		\label{fig:team-dyn-agent-rl-w-p-score}}
	\subfloat[Landmark ]{\includegraphics[width=.2\textwidth]{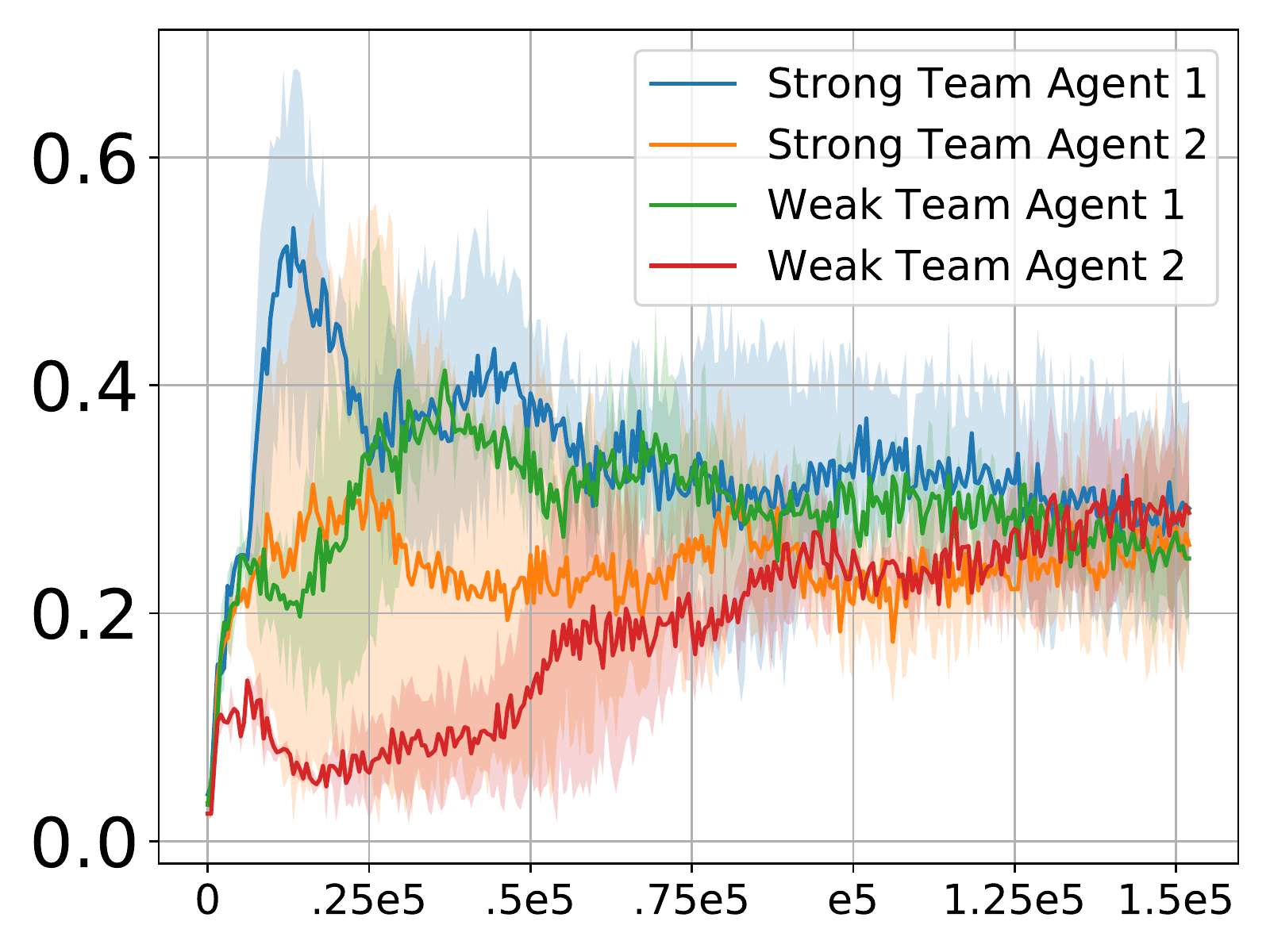}
		\label{fig:team-dyn-agent-rl-w-p-count}}	
	\subfloat[Win Policy Usage ]{\includegraphics[width=.2\textwidth]{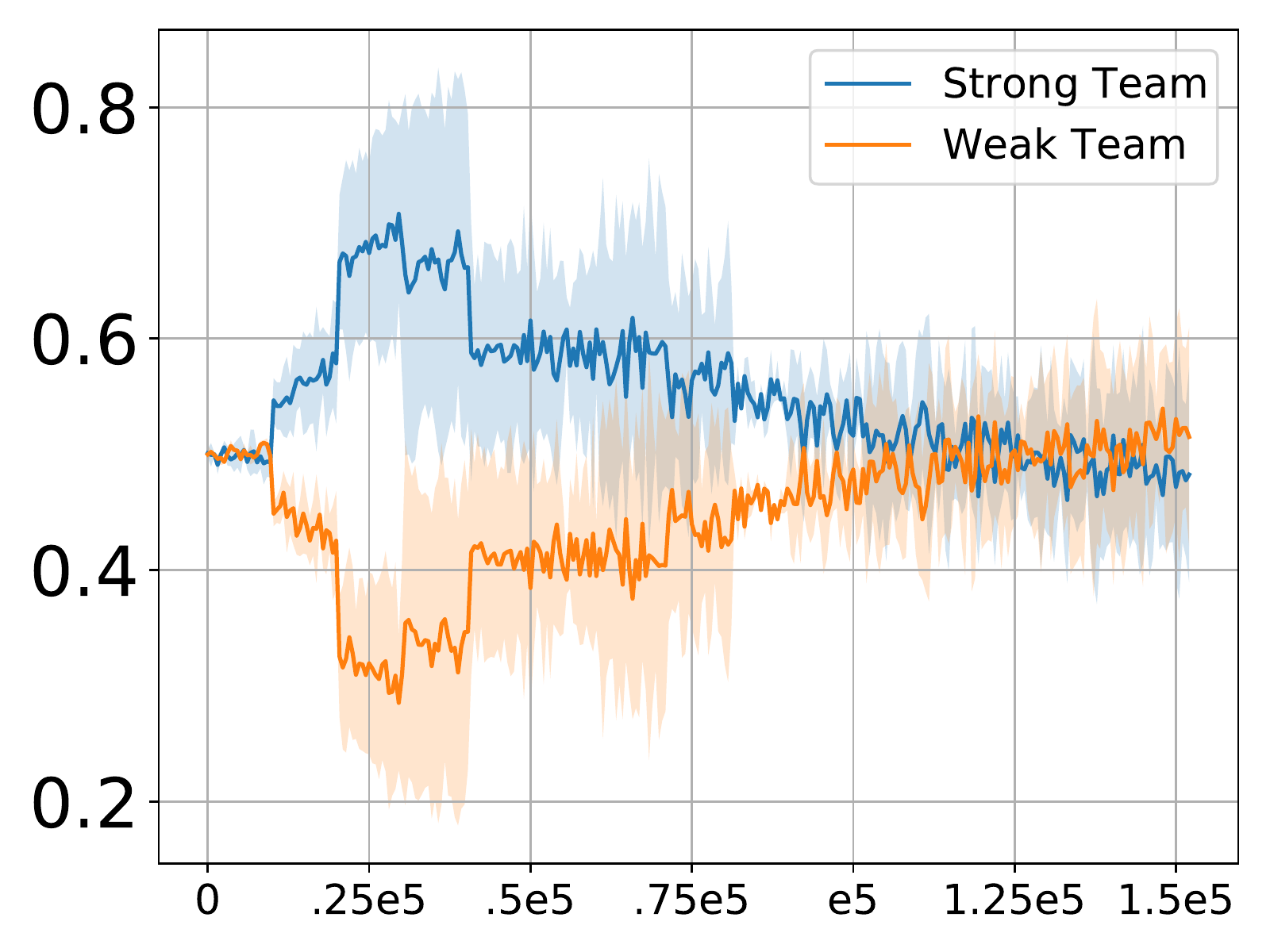}
		\label{fig:team-dyn-agent-rl-w-p-win}}		
	\subfloat[Speed ]{\includegraphics[width=.2\textwidth]{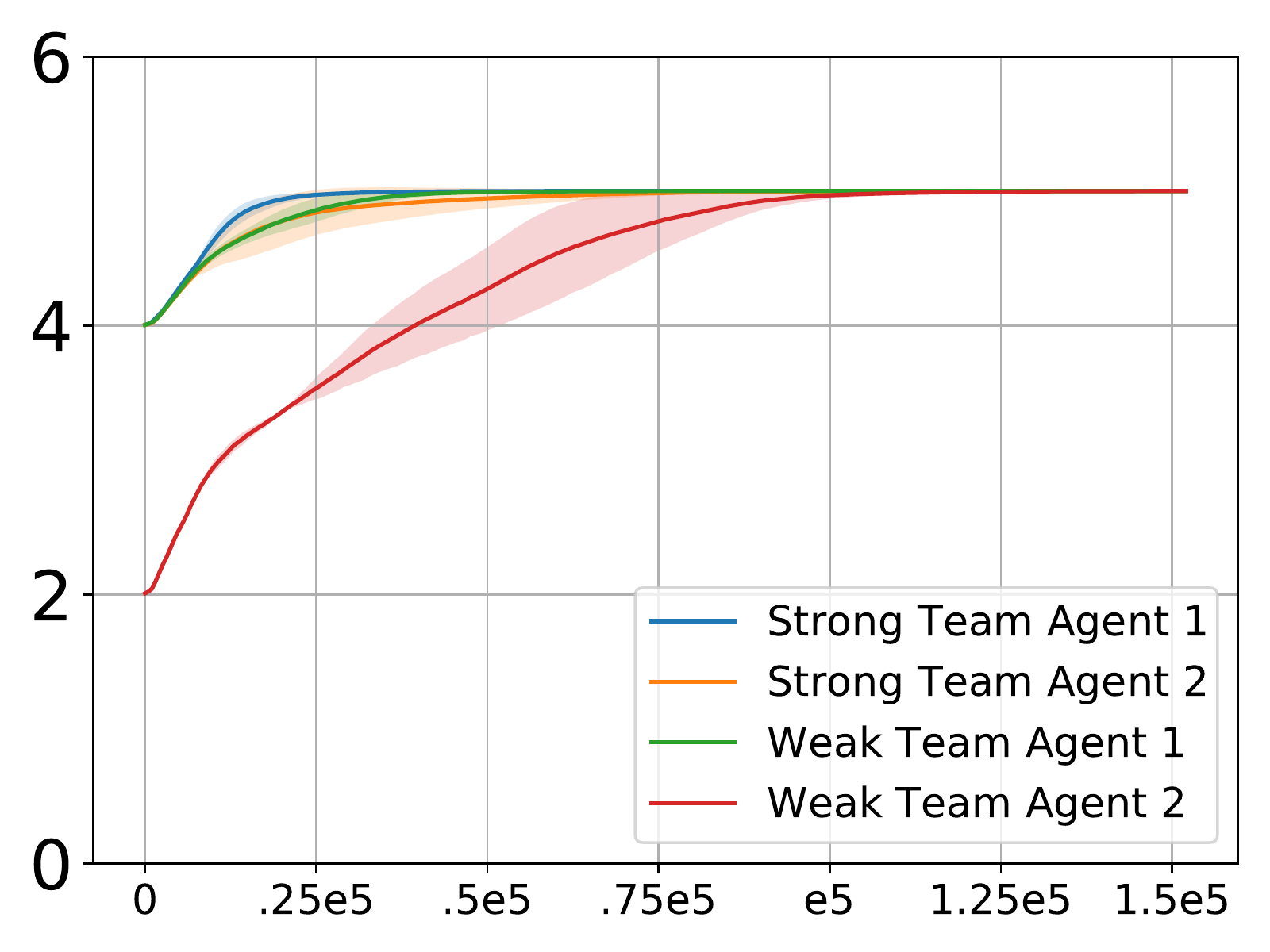}		
		\label{fig:team-dyn-agent-rl-w-p-speed}}		
	\subfloat[Incentive ]{\includegraphics[width=.2\textwidth]{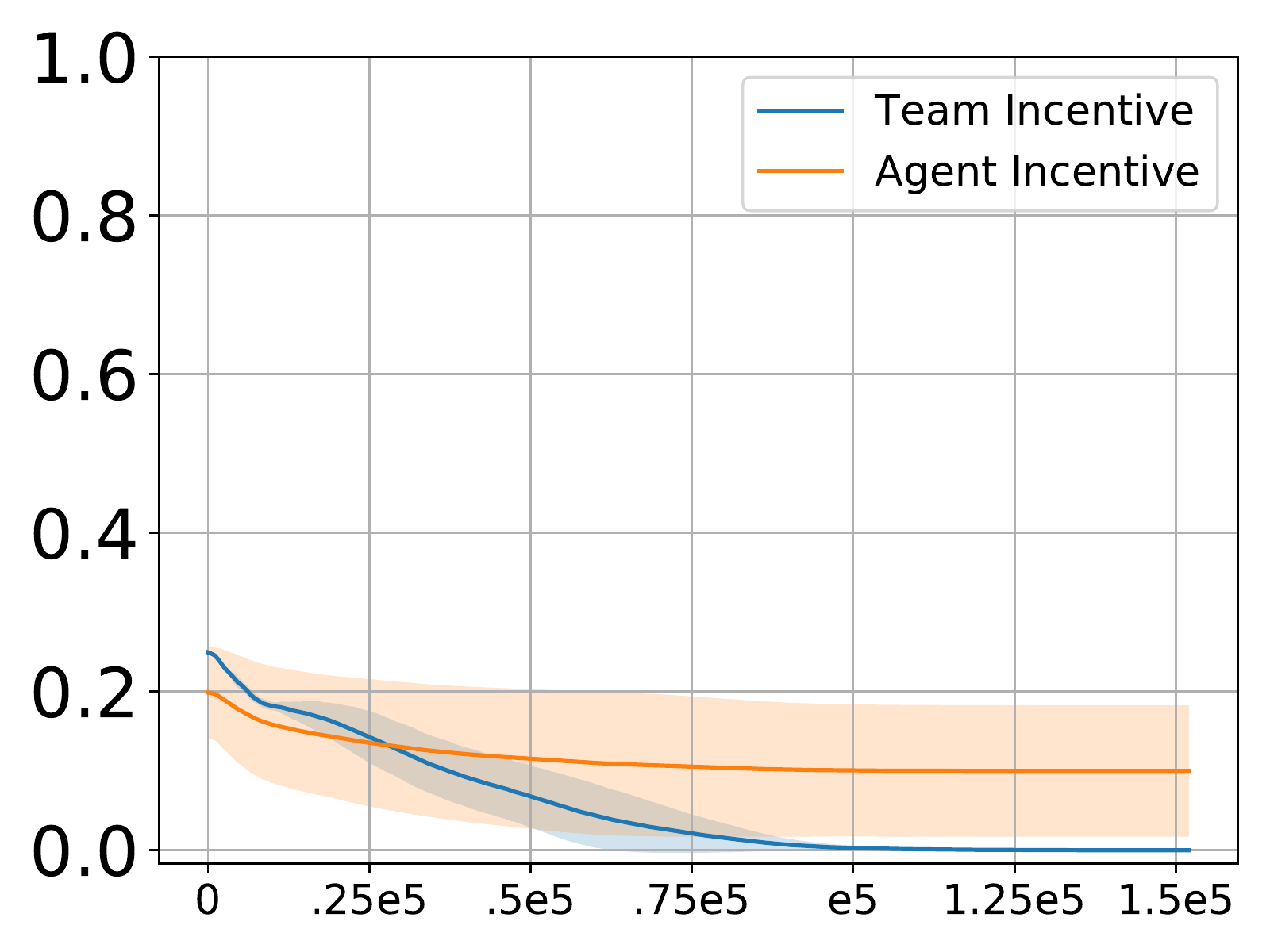}		
		\label{fig:team-dyn-agent-rl-w-p-incentive}}		
	\caption{ Agents trained in \ourgame\    game using \our{.} [Fig. (a)-(e) for Team-RL-Agent-Dynamic-Incentive scheme where Team wise reward is obtained from an RL scheme and Agent wise reward is obtained from their difference in speed. Fig. (f)-(j) for Team-Dynamic-Agent-RL-Incentive scheme where Agent wise reward is obtained from an RL scheme and Team wise reward is obtained from their difference in speed.] Average performance over $4$ different seeds has been reported (the shades signify the  $95\%$ confidence region).}
	
	\label{fig:intrinsic-rl}
\end{figure*}

\section{Dynamic Incentive Scheme}
\label{sec:dynincentive}

In the dynamic incentive scheme, the team-wise incentive ($\alpha_{\tT}$) and the agent-wise incentive ($\alpha_{\aA}$) are determined dynamically at each timestep, considering  
(1) {\bf Landmark-based Incentive} - using the difference in performance  (landmark touching count) of agents or (2) {\bf Speed-based Incentive} - using the
difference in speed (skill) of agents. 
Hence: 
\begin{equation}
\begin{gathered}
\alpha_{\tT} =   n_{\tT(strong)}-n_{\tT(weak)},\\
 ~\alpha_{\aA} = n_{\aA(strong,\tT(weak))}-n_{\aA(weak,\tT(weak))}
\label{eq:alpha2}
\end{gathered}
\end{equation}
where $n_{\tT(x)}$, $n_{\aA(x)}$ is either performance or speed (suitably normalized) of a team and an individual agent respectively. 
If any of the values of $\alpha_{\tT}$, $\alpha_{\aA}$ turns out to be negative at any instance, that value is set to zero.

\noindent{\bf Results: Landmark-based Incentive  } 
[\cref{fig:intrinsic-landmark-score}-\cref{fig:intrinsic-landmark-incentive}] {\bf - }
We observe the dynamic incentives help in {\em successfully balancing the final rewards} obtained by the teams. 
The incentive 
strengthens  the weaker agent 
 (see \cref{fig:intrinsic-landmark-count})) as landmark touching rate of the weaker player nears that of a strong player. For both teams, we also find that the fraction of times each agent pursues winning policy becomes 
similar ( \cref{fig:intrinsic-landmark-win}). The speed of the weaker player also catches up (\cref{fig:intrinsic-landmark-speed})  and the {\bf incentive needed to match the two teams disappears} (\cref{fig:intrinsic-landmark-incentive}). \\
\noindent{\bf Results: Speed-based Incentive} [\cref{fig:intrinsic-speed-score}-\cref{fig:intrinsic-speed-incentive}] {\bf  - } 
In this case, 
{\em the rewards of the two teams} (\cref{fig:intrinsic-speed-score}), unlike in \cref{fig:intrinsic-landmark-score}, 
{\em do not become equal}, 
even after 150000 episodes. We also see the  learning curve (indicated by number of landmark touched (\cref{fig:intrinsic-speed-count})) is low for the weak agent. The weaker team's fraction of winning policy is also comparatively less (\cref{fig:intrinsic-speed-win}). 
However, the speed of the weaker agent increases and reaches the maximum (\cref{fig:intrinsic-speed-speed}), albeit slower than in  \cref{fig:intrinsic-landmark-speed}.
This in turn pushes both $\alpha_{\tT}$ and $\alpha_{\aA}$ towards zero. 
Hence, although this incentive scheme increases the skill of the weaker team (player), the {\bf real capability} which is a {\bf complex combination of skill and policy learning lags}. Therefore, we conclude that  skill in many cases 
may not reveal the true relative positions in terms of performance.

Thus from these experiments, we find that the dynamic feedback of the difference in  landmark-count (performance) is the best
way to balance the output of the two teams. 
However, performance is a complex quantity to measure, moreover, 
competitors may not always immediately (if at all)  share this information. 
Therefore, the challenge lies in  achieving equivalent balancing without taking  performance information as an input 
in real-time; we employ an RL technique to estimate the performance. 

\begin{figure*}[!ht]
	\centering
	\subfloat{\includegraphics[width=.9\textwidth]{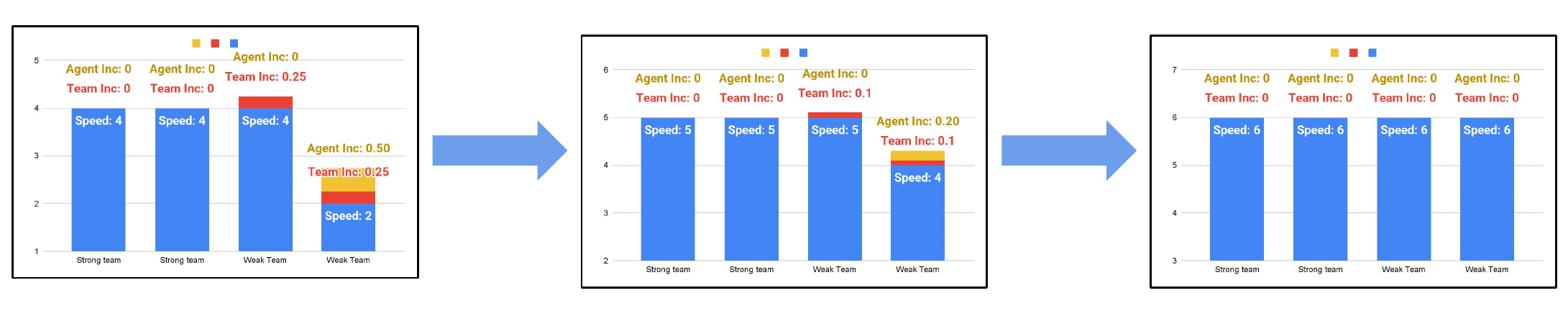}}	
	\caption{A transition diagram showing how the capabilities of the heterogenous agents evolve as the training progresses. The incentives used to close the gap between agent capabilities are diminishing as the training progresses.}
	\label{fig:transition}
\end{figure*}	
\begin{figure}[!ht]
	\centering
	\subfloat{\includegraphics[width=.5\textwidth]{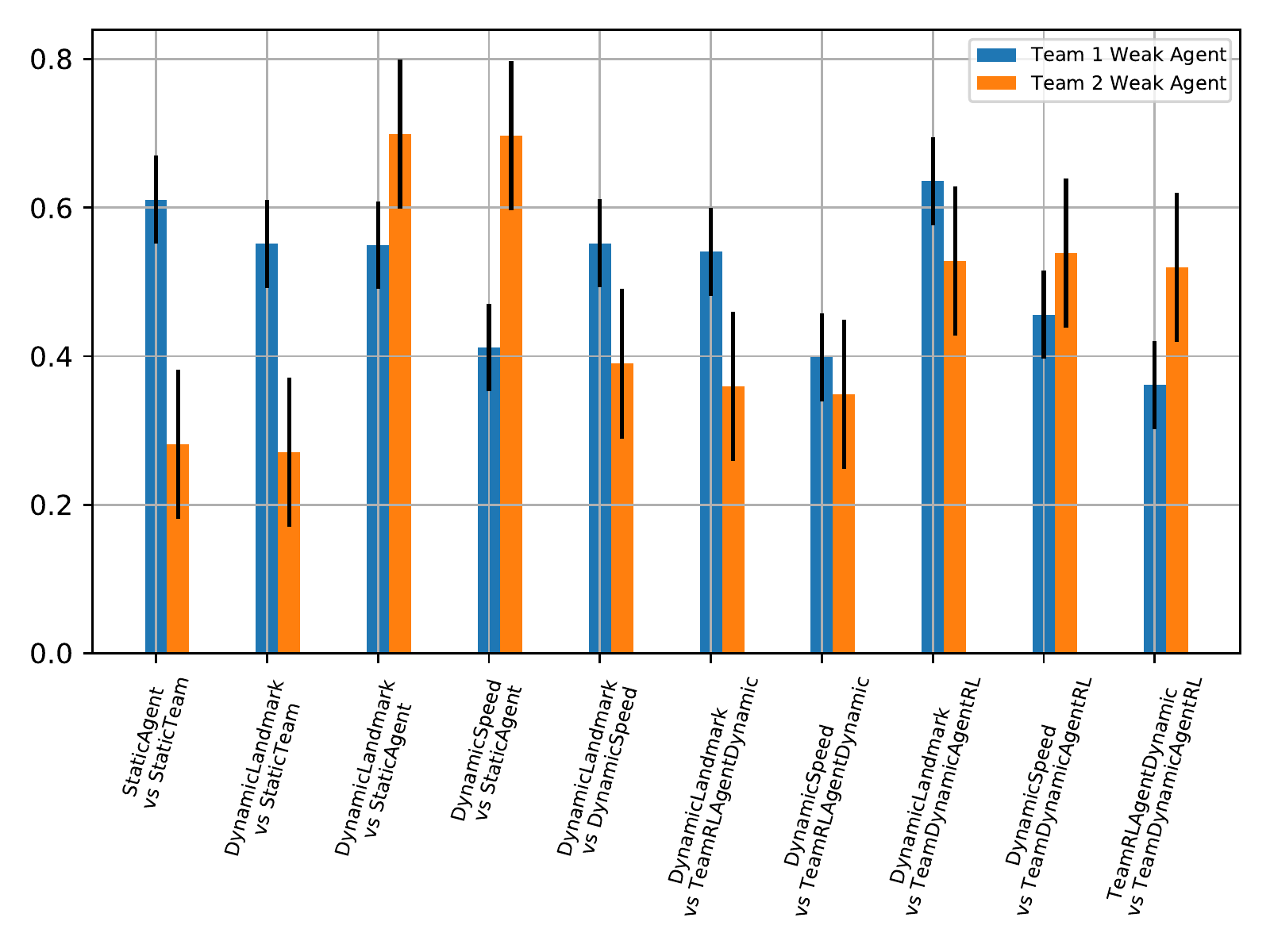}}	
	\caption{Tournament among various incentive schemes.}
	\label{fig:tournament}
\end{figure}	

\subsection{Dynamic Incentive Using Reinforcement Learning}
Here we propose two RL-based incentive mechanisms which take current speed (and not performance) of all agents as input:
(1).{\bf  [Team-RL-Agent-Dynamic]} - 
where we decide the value of $\alpha_{\tT}$ using $\pi(s)$ where $s$ is the 
current speed configuration and $\pi$ is the policy learned by training an RL model. $\alpha_{\aA}$ is computed as in \cref{eq:alpha2}. (2). 
{\bf [Team-Dynamic-Agent-RL]} -  similarly, we decide the value  of $\alpha_{\aA}$ using $\pi(s)$ and compute  $\alpha_{\tT}$ using \cref{eq:alpha2}. 

{\bf Training the RL model:}
We train the RL agent in an off-policy mode using Soft Actor-Critic \cite{haarnoja2018soft} algorithm where it uses the full training episode of \ourgame\ game setting using \our\ algorithm. 
We assume that the performance information is available during training time. 
The observation space of the RL agent consists of current speed configuration of all the agents. 
As per the need of the incentive scheme, either $\alpha_{\tT}$ or $\alpha_{\aA}$ is determined by sampling from the RL module $\pi(s)$.
The RL module obtains reward by measuring the difference in performance of the agents after applying $\alpha_{\tT	 / \aA} = \pi(s)$	for a fixed number of episodes. The intuition is that this scheme will guide the RL agent to learn the desired mapping between speed and performance.
We postpone joint training of $\alpha_{\tT}$ and $\alpha_{\aA}$ as it is extremely computationally expensive to obtain stable policies for both.

\noindent {\bf Results }: 
First we train the RL agent for 100000 episodes of underlying \our{.} We then train agents using \our\ for  $150000$ episodes 
and obtain the value of $\alpha_{\tT}$ or $\alpha_{\aA}$ using the pre-trained RL model, and summarize the results through \cref{fig:intrinsic-rl}. \\
{\bf  Team-RL-Agent-Dynamic   }[\cref{fig:team-rl-agent-dyn-w-p-score} - \cref{fig:team-rl-agent-dyn-w-p-incentive}]  {\bf -} In \cref{fig:team-rl-agent-dyn-w-p-score},
we observe that the {\bf balance of reward between the two teams gets tilted towards the weaker team } after sometime. 
We also find that asymptotically the landmark touching rate of the weaker player converges towards its stronger peers (\cref{fig:team-rl-agent-dyn-w-p-count}).
The speed of it increases (\cref{fig:team-rl-agent-dyn-w-p-speed})  and  the agent-based incentive diminishes to almost zero 
(\cref{fig:team-rl-agent-dyn-w-p-incentive}). 
However,  the team-wise incentive (\cref{fig:team-rl-agent-dyn-w-p-incentive}), doesn't vary much beyond a time, indicating that the {\em RL agent could not 
	regulate incentive value beyond initial phase }making the weaker team receive undue advantage and surpass the initially stronger team.\\
{\bf Team-Dynamic-Agent-RL}  [\cref{fig:team-dyn-agent-rl-w-p-score} - \cref{fig:team-dyn-agent-rl-w-p-incentive}] {\bf -}  We observe that the {\em reward match} between two teams is better than \cref{fig:team-rl-agent-dyn-w-p-score} -
the match {\em is almost as good as the result obtained in the case of Dynamic Landmark based incentive scheme} (\cref{fig:intrinsic-landmark-score}). 
The landmark touching rate of the weaker agent slowly increases and catches up with the stronger agents (\cref{fig:team-dyn-agent-rl-w-p-count}). 
The winning policy usage by both teams roughly become similar (\cref{fig:team-dyn-agent-rl-w-p-win})
Also, the speed of the weaker player increases steadily towards the maximum speed (\cref{fig:team-dyn-agent-rl-w-p-speed}). 
Both the incentives ($\alpha_{\tT}$, $\alpha_{\aA}$) decrease, however, the  agent incentive  persists a bit; this is because  the RL agent learns that 
mere matching of speed does not necessarily mean the attainment of capability (which was one of the learnings of speed-based incentive scheme (\cref{fig:intrinsic-speed-score} - \cref{fig:intrinsic-speed-incentive})). However, there may be some minor performance estimation error from speed  which gets reflected in smaller mismatches in score  observed in (\cref{fig:team-rl-agent-dyn-w-p-score}). 
From the results, we can thus conclude that Team-Dynamic-Agent-RL-Incentive scheme can be a good replacement for  Landmark-Based-Dynamic-Incentive scheme. Fig. ~\ref{fig:transition}
demonstrates the transition of the teams of different capabilities as the training progresses as well as the evolution of incentives used to close the gap between agent capabilities.

\section{Comparison across incentive schemes}

In this section, we explicitly compare all the incentive schemes, through direct tournaments and then comparing the performance of the weak teams, trained under different incentive schemes.

\subsection{Tournament between incentive schemes}
Here we compare various incentive schemes in a more direct way by playing them against each other in a tournament style. We train \our\ agents under various incentive schemes for $150000$ episodes for $4$ different seeds. 
Both the competing teams have one weak and one strong player; teams with  different incentive schemes are played against each other for $1000$ test episodes for $6$ different combination of competing models. \Cref{fig:tournament} reports  the landmark count per episode averaged over all $6$ experiments for the weakest member of each team, normalized by the team performance. The results confirm the following facts. 
(i) Static agent incentive scheme trains the weak member better than static-team or dynamic schemes.
(ii) Dynamic landmark scheme trains the weak agent better than dynamic-speed scheme, RL-based schemes or static schemes.
(iii) Among RL-based dynamic incentive schemes, Team-RL-Agent-Dynamic scheme trains the weak agent worse than both dynamic landmark and speed schemes. 
(iv) However, performance of the weak agent trained under team-dynamic-agent-RL is between dynamic-landmark and dynamic-speed schemes, which confirms team-dynamic-agent-rl as a good substitute of dynamic-landmark scheme.

\subsection{Comparing the variation of performance of the incentive schemes}
We remind again that our primary motivation in bringing fairness in unequal competition is to incentivize the weak team in such a way that the teams with unequal expertise should end up achieving equivalent performances. So, an incentive scheme would be optimal if it can ensure \textbf{little or no variation in performance (touching landmark)} among  all the agents. 
To find the optimal candidate among proposed incentive schemes, we check the variation and present the result in \cref{tab:fairness-table}.
For each incentive scheme, we take the last $1000$ episodes and measure the standard 
deviation~\cite{jain1984quantitative} of the landmark count of the four agents. Since all experiments have been  performed
on four seeds, the table reports the mean of the standard deviation with $95\%$ confidence interval.
Defining fair learning of unequal agents as the socially optimal goal, \cref{tab:fairness-table} shows the landmark-based dynamic scheme ( team-dynamic-agent-RL as the closest alternative) to be the best performer in this respect. This is in line with the results shown in \cref{fig:intrinsic-landmark-count,fig:team-dyn-agent-rl-w-p-count} depicting the landmark count behavior of the agents of these two best performing schemes. 
The incentivized learning process ensures that both the teams improve over time, only the incentive scheme lets the weaker team’s weaker member improve faster and catch up with her more accomplished peers.

\begin{table}[!h]
	\centering
		\begin{tabular}{|c|c|}
			\hline
			
			Methods & Landmark Count  \\\hline
			StaticTeam  &  0.446 $\pm$ 0.16    \\\hline
			StaticAgent  &  0.394 $\pm$ 0.04\\\hline
			DynamicLandmark  &  {\bf 0.248 $\pm$ 0.02}      \\\hline
			DynamicSpeed  &  0.302 $\pm$ 0.08\\\hline
			\small{Team-RL-Agent-Dynamic}  &  0.450 $\pm$ 0.11   \\\hline
			 \small{Team-Dynamic-Agent-RL}  &  {\bf 0.235 $\pm$ 0.01}\\\hline

	\end{tabular}
	\vspace{1mm}
	\caption{Standard Deviation of landmark count among the agents across incentive schemes.}
	\label{tab:fairness-table}
\end{table}

\vspace{-5mm}


\if{0} 
There are two teams (say) 1 and 2, and four agents whose index range from 1 - 4. Team 2 is the weak team with agents 3 ( strong)  and 4 (weak), while the strong team 1 has agents 1 and 2.  
Let, at time $t$, the number of landmarks touched by  agent $i$ between [$t$ - 1000, $t$] be denoted as  $N_i$. The normalized landmark count $n_i$ and the corresponding reward parameters can be accordingly calculated. 
\begin{gather}
	\vspace{-3mm}
	n_{i}^a = \frac{N_i}{\max_{i}N_i},   
	n_{i}^{t} = \sum_{j \in team_i} n_{j}^{a} 
	\label{eq:alpha1}
	\\
	\alpha_{\tT} =   n_{1}^{t}-n_{2}^{t}, ~\alpha_{\aA} = n_{3}^{a}-n_{4}^{a}
	\label{eq:alpha2}
	\vspace{-3mm}
\end{gather}
\noindent {\bf Speed-based Incentive}  $\colon$ 
Here the central controller uses the speed  information for deciding the incentive. Intuitively the team-wise reward, to be given to the weak team, is proportional to the average speed difference between the two teams and the agent-wise reward, to be given to the weak agent, is proportional to speed  difference between the two agents in the weak team. 
If we measure agents by their skill, instead of performance, $n_{agent_{i}} = \frac{S_i}{\max_{i}S_i} $, where $S_i$ denotes skill or speed instead of landmark count of agent $i$. Rest of the formulation remains same as \eqref{eq:alpha1}, \eqref{eq:alpha2}.

Using intrinsic reward according to the proposed incentives, we train \our\ for $150000$ episodes and summarize 
the results through \cref{fig:intrinsic-landmark} and  \ref{fig:intrinsic-speed}. 
We present the points scored by the teams  ( \cref{fig:intrinsic-landmark-score} ), we then look into individual members and count the number of times each of them has reached the landmark ( \ref{fig:intrinsic-landmark-count}) , 
the fraction  of epochs each of them  uses the winning policy ( \ref{fig:intrinsic-landmark-win}), 
how their individual speeds have evolved over time ( \ref{fig:intrinsic-landmark-speed}) 
and the  incentive values ($\alpha_{\tT}$, $\alpha_{\aA}$) evolving over time ( \ref{fig:intrinsic-landmark-incentive}). 
\fi


%% file: 050Conclusion.tex
\section{Conclusion}
\label{sec:conclusion}

The paper studies competition among organizations with unequal expertise and argues that certain incentive mechanisms towards the weaker members needs 
to  be formulated to ensure fair outcomes in the long run.  The entire study to devise various incentive schemes is carried out using multi-agent reinforcement 
learning framework.  However, the implementation is not straightforward, in fact, we have to devise and
 extensively test a controller assisted ensemble-based multi-agent reinforcement learning algorithm, \our\, 
which captures the importance of (winning,
losing) policy selection based on relative positions of the agents 
and facilitates the emergence of diverse roles among the agents. 
This innovative devising of controller allows us to 
formulate a setting where  the behavior of a  weaker player 
in a team is relegated more towards  a non-primary role, thus restricting its development and in turn, development of 
the entire team. 
We argue the roadblock can be removed through the introduction of targeted incentive.
We undertake a rigorous study to dynamically balance rewards and  show that there are three components
 - skill (here speed), performance (here landmark touching), and policy learning - which determine the outcome of the game and one needs to exploit their
 relationship 
to devise a practical algorithm. 

Finally, we validate the utility and effects of proposed incentive schemes on \ourgame, a simple game designed in MPE.
As a final comment, although we feel  the findings are  intriguing and reaffirm the concept of targeted subsidy, 
building a model whose outputs can be used to alleviate inequality in a  practical environment
is a non-trivial task.
Similarly, extending C-MADDPG framework towards accommodating dynamic set of policies per agent as well as allowing agents within team to simultaneously select different policies are immediate research directions to explore on the algorithmic side.